\documentclass[dvips,12pt,a4paper]{article}

\usepackage{epsfig}
\usepackage{a4}
\usepackage{amsmath}
\usepackage{latexsym}

\newcounter{figureno}
  \newlength{\absize}
\newcommand{\dd}{\mbox{{\rm d}}}
\newcommand{\half}{\textstyle\frac{1}{2}}

\renewcommand{\Re}{\mbox{Re}}
\newcommand{\thW}{\theta_{{\rm W}}}

\newcommand{\Tr}{{\rm Tr}}
\newcommand{\bslash}{\rlap/b}

\newcommand{\pslash}{\rlap/p}
\newcommand{\mslash}{\rlap/m}
\newcommand{\epsslash}{\rlap/\epsilon}
\newcommand{\ieps}{i\epsilon}
\newcommand{\alphas}{\alpha_{\rm s}}
\newcommand{\Order}{{\cal O}}
\newcommand{\calL}{{\cal L}}
\newcommand{\calQ}{{\cal Q}}
\newcommand{\Left}{{\rm L}}
\newcommand{\Right}{{\rm R}}
\def \sup{^{\vphantom{2}}}
\def\lsim{\mathrel{\rlap{\raise 2.5pt \hbox{$<$}}\lower 2.5pt
\hbox{$\sim$}}}

\newenvironment{capt}{
\phantom{mmmm}
\vspace*{10mm}
\parindent=0pt
\addtocounter{figureno}{1}

\begin{minipage}[t]{143mm}
Figure~\thefigureno.\ }{\end{minipage}
\vspace*{-5mm}}

\def\lsim{\; \raise0.3ex\hbox{$<$\kern-0.75em
      \raise-1.1ex\hbox{$\sim$}}\; }
\def\gsim{\; \raise0.3ex\hbox{$>$\kern-0.75em
      \raise-1.1ex\hbox{$\sim$}}\; }

\newcommand{\ie}{\mbox{\it i.e.}\ }
\newcommand{\etal}{\mbox{\rm et al.}, }

\begin{document}
  \thispagestyle{empty}
  \pagestyle{empty}
  \renewcommand{\thefootnote}{\fnsymbol{footnote}}
\newpage\normalsize
    \pagestyle{plain}
    \setlength{\baselineskip}{4ex}\par
    \setcounter{footnote}{0}
    \renewcommand{\thefootnote}{\arabic{footnote}}
\renewcommand{\title}[1]{%
  \begin{center}
    \LARGE #1
  \end{center}\par}
\renewcommand{\author}[1]{%
  \vspace{2ex}
  {\Large
   \begin{center}
     \setlength{\baselineskip}{3ex} #1 \par
   \end{center}}}
\renewcommand{\thefootnote}{\fnsymbol{footnote}}

\renewcommand{\thanks}[1]{\footnote{#1}}
\renewcommand{\abstract}[1]{%
  \vspace{2ex}
  \normalsize
  \begin{center}
    \centerline{\bf Abstract}\par
    \vspace{2ex}
    \parbox{\absize}{#1\setlength{\baselineskip}{2.5ex}\par}
  \end{center}}
\begin{flushright}
\texttt{CERN-TH/98-134 \\[-1mm]
hep-ph/9804389}\\[-4mm]
\end{flushright}
\vspace*{4mm}
\vfill
\title{On the phenomenology of a $Z'$ coupling only}
\title{to third-family fermions}

\author{
A.A.~Andrianov$^{a}$,
P. Osland$^{b,c}$, 
A.A.~Pankov$^{c,d}$,
N.V.~Romanenko$^{e}$
{\rm and} J. Sirkka$^{f}$}
 
\begin{center}
$^a$ Department of Theoretical Physics,
       St.-Petersburg State University,\\
       198904 St.-Petersburg, Russia \\
$^b$ Department of Physics, University of Bergen, \\
       All\'{e}gaten 55, N-5007 Bergen, Norway \\
$^c$ Theoretical Physics Division, CERN, CH 1211 Geneva 23, Switzerland \\
$^d$ Gomel Polytechnical Institute, Gomel, 246746 Belarus \\
$^e$ Petersburg Nuclear Physics Institute, Gatchina, \\
Leningrad district, 188350, Russia \\
$^f$ Department of Physics, University of Turku,
                 FIN-20500 Turku, Finland
\end{center}
\vfill
\abstract
{The phenomenology of an additional $U(1)$ neutral gauge boson
$Z'$ coupled to the third family of fermions is discussed.
One might expect such a particle to contribute to processes
where taus, $b$ and $t$ quarks are produced.
Precision data from LEP1 put severe constraints on the
mixing and heavy-boson mass.
We find that the effects of such a particle could not be observed
at hadronic colliders, be it at the Tevatron or the LHC, 
because of the QCD background.
At LEP2 and future $e^+e^-$ linear colliders, 
one could instead hope to observe such effects, 
in particular for $b\bar b$ final states.
}
\vspace*{15mm}
\vfill

\begin{flushleft}
\texttt{CERN-TH/98-134 \\[-1mm]
April 1998}
\end{flushleft}

\renewcommand{\thefootnote}{\arabic{footnote}}
\setcounter{footnote}{0}

\section{Introduction}
\setcounter{equation}{0}
We study phenomenological implications of the breaking of universality in
weak interactions. Particular attention is paid to the signatures of 
new heavy elementary vector particles 
($Z'$ bosons), which can be detected at the present
or future high-energy $e^+e^-$ and hadron colliders.
As the quarks of the third family ($t$ and $b$) are significantly 
heavier than those of the other two families, one may speculate
that such a boson may be involved in the formation of this difference, 
and thereby interact directly with the $t$ and $b$ quarks 
(as well as the $\tau$ and $\nu_\tau$)
and only much more weakly with the lighter quarks and leptons,
solely due to radiatively-created mixings.

Both the phenomenology of $Z'$-boson production and theoretical models
for its dynamics are considered.
A theoretical extension of the Standard Model adopting a new heavy $Z'$ boson
may need to include more unknown heavy fermions and scalar Higgs bosons.
Otherwise the quantum anomalies break the unitarity and the mass pattern for
heavy fermions will violate experimental bounds on deviations from the
Standard Model in the low energy data. 

There can be a variety of 
possible models depending on the strength of the vector 
and axial-vector couplings of the $Z'$ to quarks and leptons. 
Additional bosons naturally appear in Grand Unification
Theories \cite{Hewett,Lan}. The left--right-symmetric model also contains
at least one $Z'$ boson \cite{Mohapatra}.

Usually these $Z'$ bosons interact universally with the fermions
of all families; the phenomenology of such particles was
considered by several authors \cite{Hewett,Lan,Leike0,others}.
There exists a scheme in which Grand Unification symmetry breaking
leads to the formation of a so-called `leptophobic' and
`hadrophilic' $Z'$ boson, which interacts only with hadrons.
The phenomenology of such bosons was also considered 
\cite{Lan,GeorgiChiap,Barger}. A related idea has been advanced 
by Okun and collaborators, who consider a `leptonic' photon \cite{Okun}.

A scheme with the breaking of universality in weak interactions 
has been studied
by Dyatlov \cite{Dyatlov} in connection with the problem of
the fermion mass hierarchy.
It was shown that a neutral gauge boson interacting differently
with the heaviest fermions naturally leads to a realistic
mass hierarchy and Kobayashi-Maskawa mixing matrix.
The phenomenology of a $Z'$ interacting only with the quarks
of the third family was considered recently 
by Holdom \cite{Holdom},
by Frampton, Wise and Wright \cite{FWW},
and by Muller and Nandi \cite{MuNa}. 

Irrespectively of the scheme of the $Z'$-boson embedding into an extended
electroweak theory one should know the feasibility of detecting such a
particle at a particular accelerator. Various high-energy colliders
have been examined  for that purpose: 
hadronic colliders (Tevatron and the LHC),
LEP2 and future electron--positron colliders (NLC).

For $p\bar p$ and $pp$ colliders the main production mechanisms 
are the following:
$b\bar b$-quark annihilation (from the small sea contribution 
in protons) near the $Z'$ peak, quark--antiquark annihilation
into a $Z$ with conversion of the $Z$ into
a $Z'$ due to the small mixing and also gluon--gluon fusion. 
The $Z'$ boson is assumed to be detected through an excess 
of $b \bar b$ or $t \bar t$ hadronic jets. 
Other processes are suppressed by the kinematics and due to 
small coupling constants. 
Numerical estimates have been made to determine whether 
the $Z'$ signal can compete with the two-jet background 
originating from non-resonant processes. 
The conclusion is that the direct observation
of a heavy $Z'$ (heavier than the $Z$) with moderate 
coupling constants to $b$ and $t$ quarks is impossible at the $p \bar p$
Tevatron collider and unlikely even at the LHC.

The most interesting possibility is to investigate the feasibility of
a $Z'$ discovery at LEP2 \cite{YellowBook}
and the next generation of $e^+e^-$ colliders
whose design is now intensively discussed \cite{NLC}.
For these colliders the background of $\tau^+\tau^-$ and
two ($b\bar b$ or $t\bar t$) jets 
is significantly lower than for the previous types of colliders;
therefore there is a possibility of having kinematical windows 
for detecting a $Z'$ boson.

The paper is organized as follows: In Sect.~2 we set up the notation
and discuss constraints on the $Z'$ from electroweak precision data.
In Sect.~3 we consider production at hadronic colliders, and in
Sect.~4 we consider $e^+e^-$ colliders.
Section~5 contains a brief summary.
In Appendix~A we discuss the constraints on the (expanded) scalar sector 
due to anomaly cancellations, and in Appendix~B we give some formulas
on the QCD background.
\section{Model parameters and experimental input}
\setcounter{equation}{0}

We consider a model with an additional 
massive neutral $U(1)$ gauge boson $Z'$.
The Lagrangian can be written in the form:
$$ \calL=\calL_{\rm SM}+  \calL_{Z'} $$
where $\calL_{\rm SM}$ is the full Lagrangian of the Standard Model
(SM) and  $\calL_{Z'}$ is an additional term:
$$ \calL_{Z'}= \calL_{\rm YM}+ \calL_{\rm Higgs}+  
\calL_{\rm int}. $$
Here $\calL_{\rm YM}$ is the usual free Yang--Mills Lagrangian
for the $Z'$ boson, $\calL_{\rm Higgs}$ the Lagrangian
for scalar particles interacting with the $Z'$
(one may consider different choices for the Higgs sector---some 
possible Higgs structures are discussed in Appendix~A---see also
\cite{FWW,Altarelli}), and $\calL_{\rm int}$ specifies the
interactions of the additional neutral boson with fermions.
\subsection{Additional NC interactions}

We assume that the $Z'$ interacts only with fermions
of the third family. 
Then  the neutral-current
Lagrangian is of the form \cite{Lan}:
\begin{equation}
-\calL_{\rm NC}= e J^{ \mu}_{\rm em} A_{\mu}+
g_Z J^{\mu}Z_{\mu}+ g_{Z'} J'{}^{\mu}Z'_{\mu},
\end{equation}
where the first two terms are just those of the SM
and $ J'{}^{\mu}$ involves only the third-family fermions,
$t$, $b$, $\tau$ and $\nu_{\tau}$.
This form of $\calL_{\rm NC}$ corresponds to the case where
the new $U(1)$ neutral gauge particle is not mixed with the photon
\cite{Altarelli}.

In the usual notation of the SM one has $ g_Z= g/ \cos \thW$,
while $g_{Z'}$ is a phenomenologically free parameter.
However, in more complete theories involving $Z'$
(\ie GUT or LR models), the value of $g_{Z'}$ is tightly
connected to $g_Z$ \cite{Lan,Barger}:
\begin{equation}
\label{Eq:2-2}
g_{Z'}= \sqrt{\frac{5}{3}} \sin\thW \sqrt{\lambda}\, g_Z,
\end{equation}
where $\lambda$ is usually in the range $2/3$ to 1 \cite{Lan}.
This means that the maximum value is
\begin{equation}
\label{Eq:gZprime}
g_{Z'}/g_Z\approx 0.62.
\end{equation}
We shall use this constraint in most of the numerical work
that is presented.

Let us write down the Lagrangian for interactions of the $Z$ and $Z'$
with the fermions of the third family:
\begin{equation}
-\calL_{\rm ZZ'}= 
g\sup_Z Z_{\mu} 
\sum_{f=t b \tau \nu_\tau} \bar \psi_f \gamma^{\mu}
\left( v_f -a_f \gamma_5 \right) \psi_f
+g\sup_{Z'} Z'_{\mu} \sum_{f=t b \tau \nu_\tau}
\bar \psi_f \gamma^{\mu}
\left(v_f' -a_f'\gamma_5 \right) \psi_f.
\label{ZZ}
\end{equation}
The parameters $v_f$ and $a_f$ are defined in the SM as follows:
\begin{equation}
v_f= \half T_{\rm 3L}-\sin^2\thW Q, \qquad 
a_f=\half T_{\rm 3L},
\end{equation}
where $T_3$ and $Q$
are the third component of weak isospin and of electric charge,
respectively (we suppress their flavour indices).
The parameters $v_f'$ and $a_f'$ represent the chiral properties
of the $Z'$ interactions with fermions and the relative strengths of
these interactions. One of all these parameters for all fermions
can be absorbed into $g_{Z'}$, and in order to avoid this
we normalise $v_{\tau}'{}^2+ a_{\tau}'{}^2=\frac12$;
this normalization was used in \cite{Barger} when
obtaining Eq.~(\ref{Eq:2-2}).
However, in our model the quantity
$v_f'{}^2+a_f'{}^2$ may vary from fermion to fermion.
The cancellation of ABJ anomalies restricts the possible
choice of $v_f'$ and $a'_f$ and will be discussed in Appendix A.

The weak eigenstates $Z_{\mu}$ and $Z'_{\mu}$ may be mixed so that 
the mass eigenstates are:
\begin{equation}
\left(\begin{matrix}Z_1 \\ Z_2 \end{matrix}\right)=
\left(
\begin{matrix}\cos \xi  & \sin \xi \\
              -\sin \xi & \cos \xi \end{matrix}
\right)
\left(\begin{matrix} Z \\ Z' \end{matrix} \right).
\end{equation}
Then the well-known value of the SM $Z$-boson mass corresponds
in this model to the  $Z_1$ mass ($M_1$),
while the mass of the $Z_2$ boson ($M_2$) is, of course, unknown.
It is possible to rewrite Eq.~(\ref{ZZ}) in terms of the
mass eigenstates:
\begin{eqnarray}
-\calL_{\rm Z_1Z_2}&=&
g_Z \biggl[Z_{1\mu} \sum_{f=t b \tau \nu_\tau}\bar \psi_f \gamma^{\mu}
\left(v_{f1} -a_{f1}\gamma_5 \right) \psi_f \nonumber \\
& & \hspace{4mm}
+ Z_{2\mu} \sum_{f=t b \tau \nu_\tau}
\bar \psi_f\gamma^{\mu}
\left(v_{f2} -a_{f2} \gamma_5 \right) \psi_f \biggr],
\label{Z1Z2}
\end{eqnarray}
where
\begin{eqnarray}
\label{Eq:couplings}
v_{f1}&=&\cos \xi\, v_f + \frac{g_{Z'}}{g_Z} \sin \xi\, v_f', \qquad
a_{f1}= \cos \xi\, a_f + \frac{g_{Z'}}{g_Z} \sin \xi\, a_f', \nonumber \\
v_{f2}&=&\frac{g_{Z'}}{g_Z}\cos \xi\, v_f' -  \sin \xi\, v_f, \qquad
a_{f2}= \frac{g_{Z'}}{g_Z}\cos \xi\, a_f' -  \sin \xi\, a_f.
\end{eqnarray}

The introduction of mixing between the $Z$ and $Z'$ 
changes the couplings of the SM, which are 
extracted from the conventional experimental input.
In fact, in the SM one has the standard electroweak input \cite{Sch}
of $\alpha(0)$, $G_{\rm F}$ \cite{PDG}
together with $M_Z=(91.1863 \pm 0.002)\; {\rm GeV}$ \cite{Blondel}.
This input contains enough parameters for the determination
of the SM electroweak coupling constants $g$ and $g'$ of
$SU(2)_\Left$ and $U(1)_Y$, respectively, 
and the Higgs vacuum expectation value $v$.
In order to do this let us introduce the following quantities:
\begin{equation}
\mu^2\equiv\frac{\pi \alpha(0)}{\sqrt{2}G_{\rm F}} 
\approx (37.280 \; {\rm GeV})^2
\end{equation}
and \cite{Veltman77,rho-def}
\begin{equation}
\rho\equiv \frac{G_{\rm F}^{\rm NC}}{G_{\rm F}^{\rm CC}}
\end{equation}
where $G_{\rm F}^{\rm NC}$ is the four-fermion NC
Fermi coupling constant in the limit of zero momentum transfer.
At the tree level the $\rho$ parameter can be defined as
$\rho_0=M_W^2/(M_Z^2\cos^2\thW)$ and in the SM $\rho_0=1$
due to the doublet Higgs structure, while at the one-loop
level it acquires radiative corrections 
\cite{Veltman77}\footnote{For $b\bar b$ 
final states there are extra vertex corrections \cite{Altarelli}.}
\begin{equation}
\rho\equiv\frac{1}{1-\Delta \rho_T}, \qquad
\Delta \rho_T \approx\frac{3G_{\rm F}}{8 \pi^2 \sqrt{2}}\,m_t^2 
\approx 0.01.
\end{equation}

Then it is possible to express the couplings of
Eq.~(\ref{ZZ}) in terms of the above-mentioned 
input \cite{Altarelli}:
\begin{equation}
g_Z=(4 \sqrt{2} G_F M_Z^2 \rho)^{1/2},
\label{Eq.:gZ}
\end{equation}
\begin{equation}
\sin^2 \thW = \frac12-\sqrt{\frac14-\frac{\mu^2}{\rho M_Z^2}
\,\frac{\alpha(M_Z)}{\alpha(0)}}.
\label{sub}
\end{equation}
It should be stressed that the angle introduced above
differs from the conventionally defined Weinberg angle:
\begin{equation}
\sin^2 \theta_0 \cos^2 \theta_0 =
\frac{\mu^2}{ M_Z^2} \, \frac{\alpha(M_Z)}{\alpha(0)}.
\label{SW}
\end{equation}
The one-loop expression for this difference can be found in 
\cite{Peskin}.

In the extended model the mass of the weak eigenstate
$M_Z$ (\ie the matrix element of the mass operator)
is unknown: it is related to the mass
values $M_1$ and $M_2$, and to the mixing $\xi$, as
\begin{equation}
\tan^2 \xi= \frac{M_Z^2-M^2_1}{M_2^2-M_Z^2}.
\end{equation}

As a result, {\it for non-zero mixing angles there are
corrections to the Weinberg angle
$\thW$ and gauge couplings $g$ and $g'$}.
They acquire dependence, now not only from the standard
electroweak input, but {\it from the parameters of the
extended model ($\xi$, $M_2$) as well}.
This means that the results of a measurement of
other electroweak quantities may be used to restrict
these parameters.

Following \cite{Altarelli} one finds that in the extended 
model Eq.~(\ref{sub}) should be modified by
$M_Z \rightarrow M_1$, while the $\rho$ parameter acquires
additional contributions at the tree level:
\begin{itemize}
\item
from the $ZZ'$ mixing ($\Delta \rho_{\rm M}$):
\begin{equation}
\rho_{\rm M}=1+ \sin^2 \xi \left(\frac{M_2^2}{M_1^2}-1 \right)=
\frac{1+ (M_2^2/M_1^2)\tan^2 \xi }{1+\tan^2 \xi};
\label{rhom}
\end{equation}
\item
from the additional Higgs sector---related to the pattern of 
symmetry breaking ($\Delta \rho _{\rm SB}$)
(the scalar sector of this theory necessarily 
contains additional Higgs fields),
\end{itemize}
so that:
\begin{eqnarray}
\rho&=&\rho_{\rm SB} \cdot \rho_{\rm M} \cdot \rho_T
=\frac{1}{1-\Delta\rho}, \nonumber \\
\Delta \rho&=& \Delta \rho_{\rm SB} + \Delta  \rho_{\rm M} 
+ \Delta \rho_T,
\label{Del-rho}
\end{eqnarray}
where $\rho_{\rm SB}$ and $\rho_{\rm M}$ contain tree-level
corrections due to the extension of the model, whereas $\rho_T$
contains the SM radiative corrections (dominated by the top quark).
Of course, in the extended $Z'$ model there exist additional radiative
corrections to the $\rho$ parameter, which can even be rather
large since there is no decoupling of the heavy particles for
these corrections, but we restrict ourselves by considering
only the tree-level effects of the non-standard particles. 
  
If the scalar sector of the theory contains only doublets and
singlets of the $SU(2)_\Left$ group one has $\Delta \rho_{\rm SB}=0$.
In further considerations we shall use this assumption by default.
\subsection{Existing bounds on the $Z'$ mass}
A $Z'$ would contribute with full strength to the decay
of $\Upsilon$ states to $\tau^+\tau^-$, suppressed only by the
$Z'$ propagator. The moderate agreement of the branching ratio
\cite{PDG}
\begin{equation}
\mbox{BR}(\tau)\equiv\frac{\Gamma(\Upsilon(1S)\to\tau^+\tau^-)}
     {\Gamma(\Upsilon(1S)\to \mbox{all})}
=0.0297\pm0.0035,
\end{equation}
with 
\begin{equation}
\mbox{BR}(e)=0.0252\pm0.0017, \qquad
\mbox{BR}(\mu)=0.0248\pm0.0007,
\end{equation}
implies that the $Z'$ mass cannot be too low.

Let us estimate this lower bound.
Since the $\Upsilon(1S)$ is a natural-parity state, only the vector
part of the hadronic current contributes to its decay.
Hence, no restriction can be obtained for the pure axial
$Z'b\bar b$ vertex without $ZZ'$ mixing (such mixing effects are
small, and will be omitted from the discussion of $\Upsilon(1S)$ decay).
For couplings involving the vector current, one may obtain
a lower bound on the $Z'$ mass:
$M_{Z'}>50$~GeV for vector--vector chiralities (in the $Z'b\bar b$
and $Z'\tau^+\tau^-$ vertices),
$M_{Z'}>22$~GeV for the vector--axial-vector case, and
$M_{Z'}>35$~GeV for the LL, RR, LR and RL cases.

Another kind of restriction can be obtained from the $\rho$ parameter,
as discussed above.
If one combines the LEP results with the $M_W/M_Z$ value, it is possible
to conclude that \cite{Altarelli97}
\begin{equation}
\label{rho-bounds}
2\cdot10^{-3}<\Delta\rho<8\cdot10^{-3} \qquad
\mbox{for}\quad 150~\mbox{GeV}<m_t<200~\mbox{GeV}
\end{equation}
and $70~\mbox{GeV}<m_H<1000~\mbox{GeV}$.
Under the assumption that $\Delta\rho_{\rm SB}=0$,
this leads to
\begin{equation}
1\lsim \rho_{\rm M}\lsim 1.005.
\end{equation}
The lower bound on $\rho_{\rm M}$ implies that $M_2>M_1$.

However, one should keep in mind that these constraints
are valid only under the assumption that $\rho_{\rm SB}=1$,
which requires a certain simple structure of the Higgs sector.
For a more general scalar sector one has \cite{KenLyn}
\begin{equation}
\rho_{\rm SB}=1
+\frac{\sum_{{\rm all }H}
\langle H_0^\dagger|(\vec T^2-3T_3^2)|H_0\rangle_{\rm vac}}
{2\sum_{{\rm all }H}
\langle H_0^\dagger|T_3^2|H_0\rangle_{\rm vac}}\, .
\end{equation}
For Higgs fields with isospin higher than 1/2, $\rho_{\rm SB}$ may
be less than unity and the above-mentioned constraints are removed.

\section{Hadronic production}
\setcounter{equation}{0}
The hypothetical $Z'$ boson could be produced in proton--proton
(antiproton) collisions 
through direct Drell--Yan-like coupling of $b$ and $\bar b$ sea quarks, 
through gluon--gluon fusion, with the gluons
coupled to $b$ or $t$ quark loops, 
as well as through $ZZ$ fusion and $WW$ fusion.
We shall consider these mechanisms in turn in the following,
focusing first on the prospects for producing such vector bosons
at the Fermilab Tevatron, and then turn to the LHC in Sect.~3.4.
In this section we consider the limit of no $ZZ'$ mixing,
which implies $M_{Z'}=M_2$.
Also, we shall here disregard the constraint (\ref{Eq:gZprime}),
letting $Z'$ couplings be of $\Order(1)$.
\subsection{Direct production by $b\bar b$ annihilation}
The vector boson $Z'$ may be produced by the direct Drell--Yan
mechanism, \ie annihilation of a $b$ quark from one proton
(or antiproton) and a $\bar b$ from the other.
The matrix element is proportional to
\begin{equation}
M=g_{Z'}^2\bigl[\bar v(p_2)\gamma^\mu(v_b'-a_b'\gamma_5)u(p_1)\bigr]
\frac{-g_{\mu\nu}+k_\mu k_\nu/M_{Z'}^2}{k^2-M_{Z'}^2+iM_{Z'}\Gamma_{Z'}}
\bigl[\bar u(q)\gamma^\nu(v_{q}'-a_{q}'\gamma_5)v(\bar q)\bigr],
\end{equation}
where $p_1$ and $p_2$ are the relevant parton momenta
of the incident proton and antiproton, with $k=p_1+p_2$,
and $q$ and $\bar q$ the final-state quark and antiquark momenta.
The corresponding polarisation-averaged square becomes
\begin{eqnarray}
\label{Eq:b-bar-b}
\overline{|M|^2}
&=&\frac{4g_{Z'}^4}{(k^2-M_{Z'}^2)^2+(M_{Z'}\Gamma_{Z'})^2}
\bigl\{X_0 +m_b^2 X_b + m_q^2 X_q + m_b^2 m_q^2 X_{bq} \nonumber \\
& & \phantom{\frac{4g_{Z'}^4}{(k^2-M_{Z'}^2)^2+(M_{Z'}\Gamma_{Z'})^2}}
+ m_b^4 X_{bb} + m_q^4 X_{qq}
\bigr\},
\end{eqnarray}
where $m_q$ ($=m_b$ or $m_t$) denotes the mass of the final-state
quarks and
\begin{eqnarray}
X_0&=&\half(v_b'{}^2+a_b'{}^2)(v_{q}'{}^2+a_{q}'{}^2)(\hat t^2+\hat u^2)
+2v_b' a_b' v_{q}' a_{q}' \hat s (\hat t-\hat u), \nonumber \\
X_b&=&(v_{q}'{}^2+a_{q}'{}^2)[3(v_b'{}^2+a_b'{}^2)(\hat s+2\hat t)
      +2v_b'{}^2\, \hat s] 
      +12v_b' a_b' v_{q}' a_{q}' \hat s, \nonumber \\
X_q&=&(v_b'{}^2+a_b'{}^2)[3(v_{q}'{}^2+a_{q}'{}^2)(\hat s+2\hat t)
      +2v_{q}'{}^2\, \hat s] 
      +12v_b' a_b' v_{q}' a_{q}'\, \hat s, \nonumber \\
X_{bq}&=&4\{v_b'{}^2\,v_{q}'{}^2+a_b'{}^2a_{q}'{}^2
         [3-2(\hat s/M_{Z'}^2)+(\hat s/M_{Z'}^2)^2]\}, \nonumber \\
X_{bb}&=& X_{qq}=2(v_b'{}^2+a_b'{}^2)(v_{q}'{}^2+a_{q}'{}^2),
\end{eqnarray}
with $\hat s=k^2$, $\hat t=(p_1-q)^2$, and $\hat u=(p_1-\bar q)^2$.

For the decay of the $Z'$ (of mass $M_{Z'}$) to a fermion--antifermion pair
(of mass $m_f$), we find the partial decay width at the tree level
\begin{equation}
\Gamma_f=\frac{g^2_{Z'}\,M_{Z'}}{12\pi}\sqrt{1-(2m_f/M_{Z'})^2}
\left[v_f'{}^2\left(1+\frac{2m_f^2}{M_{Z'}^2}\right)
     +a_f'{}^2\left(1-\frac{4m_f^2}{M_{Z'}^2}\right)\right]\, ,
\end{equation}
where $v_f'$ and $a_f'$ denote the vector and axial couplings,
respectively.
Thus, the total fermionic width would be given by
\begin{equation}
\Gamma_{Z'}=\Gamma_b+\Gamma_t+\Gamma_\tau+\Gamma_{\nu_\tau}\, .
\end{equation}

The inclusive cross section may be expressed in terms of a convolution
integral over $b$-quark distribution functions:
\begin{equation}
\label{Eq:dsig-dEperp}
\frac{\dd\sigma}{\dd E_\perp}
=\int\dd x_1\int\dd x_2\,
\frac{\dd^3\hat\sigma}{\dd E_\perp\dd x_1\dd x_2}\, 
f_1^b(x_1) f_2^{\bar b}(x_2),
\end{equation}
with $E_\perp$ the transverse energy of the jets,
and $f_i^q(x)$ the $b$ ($\bar b$)-quark distribution functions.
The elementary cross section is given by
\begin{equation}
\label{Eq:dsig-dx1dx2}
\frac{\dd^8\hat\sigma}{\dd x_1\dd x_2}
=(2\pi)^4\delta^{(4)}(p_1+p_2-q-\bar q)
\frac{1}{4E_1E_2 v_{\rm rel}} \overline{|M|^2}
\frac{\dd^3q}{(2\pi)^32E_q} \frac{\dd^3\bar q}{(2\pi)^32E_{\bar q}}\, ,
\end{equation}
with $\overline{|M|^2}$ the square of the matrix element in
Eq.~(\ref{Eq:b-bar-b}),
properly averaged and summed over polarisations, and
$4E_1E_2 v_{\rm rel}=2(p_1+p_2)^2\equiv 2\hat s$.
Straightforward kinematical considerations lead to:
\begin{equation}
\frac{\dd^3\hat\sigma}{\dd E_\perp\dd x_1\dd x_2}
=\frac{1}{2\pi}\, \frac{1}{8\hat s^{3/2}}\,
\overline{|M|^2}
\frac{E_\perp}{\sqrt{E_q^2-E_\perp^2}\cosh y -E_q\sinh y}\, ,
\end{equation}
where $y$ is the rapidity and
\begin{equation}
\left.
\begin{array}{l}
E_q \\ E_{\bar q}
\end{array}
\right\}=\half \sqrt{\hat s}
\left[\cosh y \pm\sinh y\sqrt{1-4E_\perp^2/\hat s}\right] \, .
\end{equation}

It is convenient to replace the integration over $x_1$ and $x_2$ by
one over rapidity and $\hat s$, the invariant mass squared of the $Z'$.
With $\tau\equiv \hat s/s$, where $s$ is the squared c.m.\ energy
of the proton--antiproton system,
we have $x_1=\sqrt{\tau}\,e^y$
and $x_2=\sqrt{\tau}\,e^{-y}$, and obtain
\begin{equation}
\label{Eq:dsig-b-dEperp}
\frac{\dd\sigma}{\dd E_\perp}
=\frac{1}{2\pi}\, \frac{1}{8s}
\int\frac{\dd \hat s}{\hat s\sqrt{\hat s}}\,
\int\dd y\, \overline{|M|^2}\,
f_1^b(x_1) f_2^{\bar b}(x_2) \,
\frac{E_\perp}{\sqrt{E_q^2-E_\perp^2}\cosh y -E_q\sinh y}\, ,
\end{equation}
where $\overline{|M|^2}$ is obtained from Eq.~(\ref{Eq:b-bar-b}).

Resulting cross sections are given in Fig.~1 for the Fermilab
energy, $\sqrt{s}=1.8$~TeV, for three values of the mass,
$M_{Z'}=$~100, 200 and 400~GeV.
We here disregard the constraint (\ref{Eq:gZprime}), taking the more
`unbiased' view that $g_{Z'}a'$ and $g_{Z'}v'$ are of $\Order(1)$.
(If we adopt Eq.~(\ref{Eq:gZprime}), the cross section would
drop by about two orders of magnitude.)
For the distribution of $b$ ($\bar b$) quarks in the incident
protons and antiprotons, we use standard values \cite{Plothow, GRV}.
For comparison, we also show the dominant QCD contributions
[cf.\ Appendix~B] and data (summed over all flavours) \cite{CDF}.
The direct $b\bar b$ production through a $Z'$ is seen to be below 
the QCD rate by 3--4 orders of magnitude, even for a `light' $Z'$.

\subsection{Gluon fusion}
In the collisions of protons and antiprotons at Fermilab
(or protons at LHC), gluon fusion may, via a suitable quark triangle
diagram, lead to production of such $Z'$ bosons.
The inclusive cross section may be expressed in terms of a convolution
integral over gluon distribution functions, similar to 
Eq.~(\ref{Eq:dsig-dEperp}):
\begin{equation}
\frac{\dd\sigma}{\dd E_\perp}
=\int\dd x_1\int\dd x_2\,
\frac{\dd^3\hat\sigma}{\dd E_\perp\dd x_1\dd x_2}\, 
f_1^g(x_1) f_2^g(x_2)
\end{equation}
with $f_i^g(x)$ gluon distribution functions.
The elementary cross section is given by an expression similar
to Eq.~(\ref{Eq:dsig-dx1dx2}), where
the amplitude for the production of $q\bar q$ ($b \bar b$ or $t \bar t$)
final states may be expressed as
\begin{equation}
M=V^{ab}_{\mu\alpha\beta}\epsilon^\alpha(1)\epsilon^\beta(2)\,
g_{Z'}\,
\frac{-g^{\mu\nu}+k^\mu k^\nu/M_{Z'}^2}{k^2-M_{Z'}^2+iM_{Z'}\Gamma_{Z'}} \,
[\bar u(q)\gamma_\nu(v_q'-a_q'\gamma_5)v(\bar q)]\, .
\end{equation}
The factor $V^{ab}_{\mu\alpha\beta}$ describes the $Z'gg$ vertex
($a$ and $b$ are colour indices), which we take from 
the corresponding expression for the $Zgg$ vertex, given by
Kniehl and K\"uhn \cite{KK}.

By current conservation,
$k^\nu[\bar u(q)\gamma_\nu v(\bar q)]=0$.
Furthermore,
for on-shell gluons, the vertex function simplifies considerably.
Averaging over gluon polarisations, and summing over $q$ and $\bar q$
polarisations, we obtain
\begin{equation}
\label{Eq-glue-M2}
\overline{|M|^2}
=4a_Q'{}^2g_{Z'}^4\left(\frac{\alphas}{\pi}\right)^2 \delta^{ab}|B_Q|^2
\left(\frac{\hat s}{M_{Z'}^2}\right)^2
a_q'{}^2 m_q^2 \, \hat s \,.
\end{equation}
Here, $a_Q'$ describes the axial coupling of $Z'$ 
to the quark field of the triangle diagram, normalised as in Sect.~2.
(If we were to replace the $Z'$ by the ordinary $Z$,
we would get $a_Q'g_{Z'}\to eT_3/(2\sin\thW\cos\thW)$, 
with $T_3=\pm\half$ the quark isospin.)
The coefficient $B_Q$ is given in Ref.~\cite{KK}.
For a triangle diagram of quark flavour $Q$, it is
\begin{equation}
B_Q=\frac{\hat s}{2\lambda}\left(1-2m_Q^2 C_0\right),
\end{equation}
with $\lambda=\lambda(s_1,s_2,\hat s)=\hat s^2$ the K\"allen function
(for on-shell gluons, $s_1=s_2=0$), 
$m_Q$ the mass of the loop quark, and $C_0$ given as
\begin{eqnarray}
C_0&=&
\frac{1}{2\hat s}(2\phi-\pi)^2, \qquad \sin\phi=\sqrt{4m_Q^2/\hat s-1}, \qquad
\hat s<4m_Q^2,  \\
C_0&=&
-\frac{1}{2\hat s}\left(\log\frac{1+\hat a}{1-\hat a}-i\pi\right)^2, \qquad
\hat a=\sqrt{1-4m_Q^2/\hat s}<1, \qquad 4m_Q^2<\hat s, \nonumber
\end{eqnarray}
depending on whether $\hat s$ is below or above the threshold associated
with the loop quark. We consider the contributions from both
$b$ and $t$ quarks to this loop:
\begin{equation}
\label{Eq:inter-b-t}
a'_Q{}^2|B_Q|^2\to |a'_bB_b+a'_tB_t|^2.
\end{equation}

The result (\ref{Eq-glue-M2}) is proportional to $a_q'{}^2m_q^2$.
According to Furry's theorem, the vector part of the triangle diagram
cancels, and the remaining axial anomaly is proportional to the 
final-state quark mass, here denoted by $m_q$.
At moderate energies, the relevant final states are $b\bar b$,
thus $m_q=m_b$.
If the energy is high enough, there is a similar contribution
for $t\bar t$ final states.

Replacing the integrations over $x_1$ and $x_2$ by integrations
over rapidity and $\hat s$, the invariant mass squared of the $Z'$,
one obtains an expression similar to Eq.~(\ref{Eq:dsig-b-dEperp}),
where the $b$ and $\bar b$ distribution functions should be replaced
by gluon distribution functions, and
where $\overline{|M|^2}$ also should contain a factor $1/8$ from colour
matching of the two gluons.

Numerical values are obtained for these cross sections, using
standard gluon distribution functions \cite{Plothow, GRV}.
The resulting cross sections are for the Tevatron energy shown in Fig.~2.
We have arbitrarily taken the axial couplings to $b$ and $t$ quarks
to be the same.
The cross section is remarkably small, even for moderately
low masses, $M_{Z'}\simeq\Order(M_Z)$.
The resonant structure is due to interference between the contributions
of $b$- and $t$-quark triangle diagrams, cf.\ Eq.~(\ref{Eq:inter-b-t}).
This interference is illustrated in Fig.~3, 
where for $M_{Z'}=200$~GeV we compare three
cases: (1) the $bbZ'$ and $ttZ'$ axial couplings being the same (solid),
(2) opposite (dashed), and (3) the $ttZ'$ axial coupling being zero.
It appears that no variation of the chirality of these couplings
can make the cross section comparable with the QCD background.
\subsection{Fusion of $ZZ$ or $WW$ bosons}
The fusion of two weak gauge bosons, like that of two gluons,
can also lead to $Z'$ production.
This could proceed through mixing, or via a triangle loop.
A crude estimate for the magnitude of this rate
may be obtained in the Weizs\"acker--Williams approximation.
The resulting cross section is found to be several orders of magnitude
below the Drell--Yan rate.
\subsection{Production at the LHC}
At the LHC the sea-quark and gluon distributions are much less suppressed
than at the `lower' energy of the Tevatron, so it is of considerable 
interest to
see if the $Z'$ production can compete with the QCD background.
We show in Figs.~4 and 5 the cross sections for $b \bar b$ 
Drell--Yan-type production and gluon fusion, respectively.
As at lower energies, it is the Drell--Yan-type production that
dominates the gluon-fusion mechanism and,
relative to the QCD rate, the $Z'$ production is now `only'
suppressed by about two orders of magnitude.
One must conclude that it would be extremely hard to discover 
such vector bosons in hadronic collisions. 

\section{Electron--positron annihilation}
\setcounter{equation}{0}
In $e^+e^-$ collisions the $Z'$ boson could be produced
directly via mixing with the $Z$ boson; the following channels
with two-particle final states are sensitive to this mixing:
\begin{equation} 
e^+e^- \rightarrow b \bar{b}; \;\; e^+e^- \rightarrow t \bar{t}
; \;\; e^+e^- \rightarrow \tau \bar{\tau}.
\end{equation}
The  first of these was discussed in \cite{Barger,FWW}.
In the following subsection we are going to discuss all these processes
and calculate the relative deviation of the cross sections
from their values in the SM for different values of the mixing angle
$\xi$ and the $Z_2$ mass $M_2$.

For the processes with more than two particles in the final state we
will consider the process 
$e^+e^- \rightarrow b \bar{b} \nu_{e} \bar\nu_{e}$ 
as one of the most important for LEP2 and future colliders.
\subsection{Two-fermion final states}

We first consider the processes 
$e^+e^-\to b\bar b$, $t \bar t$, $\tau^+\tau^-$ as the simplest processes 
at electron--positron colliders involving fermions of the third
family, where the effect of $Z'$ exchange could be observable.
Three Feynman diagrams describe this process, with the exchange of
photons, $Z_1$ and $Z_2$.
Deviations from the SM will occur only in the case where the mixing 
angle is non-zero.
(Without tree-level mixing, it will arise only through loop effects.)
The effect of $ZZ'$ mixing changes the couplings of the $Z$ boson
and gives rise to $Z_2$ exchange.
The coupling between the electron and the $Z'$ is proportional
to $\sin\xi$, which by assumption is small. On the other hand,
this mechanism has the advantage that there is no suppression
by the $Z'$ propagator and the effect might thus be observable.

The cross section is given by the expression (for $m_f\ll\sqrt{s}$):
\begin{equation}
\sigma_{f\bar f}=\frac{4\pi\alpha^2}{3s}\, F_1,
\end{equation}
where
\begin{eqnarray}
\label{Eq:F1}
F_1
&=&Q_e^2Q_f^2 +2Q_e \bar v_{e1} Q_f \bar v_{f1}\Re \chi\sup_1
+(\bar v_{e1}^2+\bar a_{e1}^2)(\bar v_{f1}^2+\bar a_{f1}^2)
|\chi\sup_1|^2 \nonumber \\[2mm]
& & +2Q_e \bar v_{e2} Q_f \bar v_{f2}\Re \chi\sup_2
+(\bar v_{e2}^2+\bar a_{e2}^2)(\bar v_{f2}^2+\bar a_{f2}^2)
|\chi\sup_2|^2 \nonumber \\[2mm]
& & +2(\bar v_{e1} \bar v_{e2} +\bar a_{e1} \bar a_{e2})
      (\bar v_{f1} \bar v_{f2} +\bar a_{f1} \bar a_{f2}) 
\Re \chi\sup_1\chi\sup_2{}^* \,,
\end{eqnarray}
with
\begin{equation}
\chi\sup_i=\frac{s}{s-M_i^2+iM_i\Gamma_i},
\end{equation}
and [cf.\ Eq.~(\ref{Eq:couplings})]
\begin{equation}
\bar v_i=\frac{g_Z}{e}\, v_i, \qquad 
\bar a_i=\frac{g_Z}{e}\, a_i\, .
\end{equation}
The ratio $g_Z/e$ should be extracted from the standard 
electroweak input discussed in Sect.~2, and include the `running'
dependence from the $e^+e^-$ energy $\sqrt{s}$.

For $\sqrt{s}=M_1$ one can use the following \cite{Altarelli, Sch}
\begin{equation}
\label{Eq:gZe}
\frac{g_Z}{e}(M_1^2)
=\frac{1}{\sin\thW \cos\thW}\bigg|_{\sqrt{s}=M_Z}
=\left[\frac{M_1^2\rho\alpha(0)}{\mu^2\alpha(M_1)}\right]^{1/2}.
\end{equation}
For $\sqrt{s}>M_1$ we will start with (\ref{Eq:gZe}) and then
use the solutions of the one-loop
massless renormalization-group equations for the $SU(2)$ and $U(1)$
running couplings, $g$ and $g'$:
\begin{eqnarray}
g^2(s)&=&
g^2(M_1^2)
\left(1+\frac{g^2(M_1^2)}{16\pi^2}\,\frac{10}{3}\log\frac{s}{M_1^2}
\right)^{-1},
\nonumber \\
g'{}^2(s)&=&
g'{}^2(M_1^2)
\left(1-\frac{g'{}^2(M_1^2)}{16\pi^2}\,\frac{20}{3}\log\frac{s}{M_1^2}
\right)^{-1},
\label{Eq:gs}
\end{eqnarray}
which are related to $e$ and $\sin\thW$ as
\begin{equation}
e^2=\frac{g^2g'{}^2}{g^2+g'{}^2}, \qquad
\sin^2\thW=\frac{g'{}^2}{g^2+g'{}^2}.
\end{equation}
It should be emphasised that all the above-mentioned couplings depend,
for the model under consideration,  on the mixing angle $\xi$
and the mass $M_2$ through the $\rho$ parameter,
cf.\ Eqs.~(\ref{Eq:gZe}) and (\ref{Eq:gs}).

No deviation from the SM has been observed at LEP.
One may thus obtain bounds on the model parameters $\xi$ and $M_2$,
taking into account available data on the $Z^0$ peak.
For LEP2 and the NLC we take a rather conservatively anticipated
precision.
The sensitivity of observables, e.g.\ of the total cross section
$\sigma_{f\bar f}$, has been assessed numerically by defining
a $\chi^2$ function as follows:
\begin{equation}
\label{Eq:chisq}
\chi^2
=\left(\frac{\Delta\sigma_{f\bar f}}{\delta\sigma_{f\bar f}}\right)^2,
\end{equation}
where $\Delta\sigma_{f\bar f}=\sigma_{f\bar f}-\sigma^{\rm SM}_{f\bar f}$ 
and the uncertainty $\delta\sigma_{f\bar f}$ is the statistical one. 
As a criterion to derive allowed regions for
the coupling constants if no deviations
from the SM were observed, and in this way to assess the sensitivity
to the parameters $\xi$ and $M_2$,
we impose that $\chi^2<\chi^2_{\rm crit}$, where $\chi^2_{\rm crit}$
is a number that specifies the desired level of significance.

At the $Z^0$ peak, the most sensitive quantity is the forward--backward
asymmetry. The resulting allowed bounds on $\xi$ and $M_2$, at the 95\%
C.L., are given for $b\bar b$ and $\tau^+\tau^-$ production
in Figs.~6--8, for different assumed chiralities of the coupling to
the $Z'$ (vector, axial, left, right).
The relative gauge coupling is chosen according to \cite{Lan} as
$g_{Z'}/g_Z\simeq0.62$.
These bounds were obtained from the data reported in \cite{LEP1},
by means of the program ZEFIT, which has to be used along with ZFITTER
\cite{zfitter}.

In these same figures, we also present bounds corresponding
to conservatively assumed cross-section precisions of 5\% and 10\% 
at LEP2 and a linear collider operating at 500~GeV (labelled NLC).
These cross sections were calculated by means of
the CompHEP \cite{CompHEP} program.
The qualitative difference between the LEP1 and the LEP2 (or NLC)
contours is due to the following.
At LEP1 (on resonance) there are three effects: modification of the
couplings due to mixing, modification of $\sin^2\thW$, and a shift
of $\rho$ from the SM value.
The corresponding bounds are smooth curves.
At higher energies (LEP2 and NLC) there is an additional contribution
mediated by $Z_2$ exchange.
At relatively low $M_2$ values, the $Z_2$-exchange contribution
dominates the deviation from the SM, whereas at higher $M_2$ 
this contribution becomes less important.
The complicated shape of the contours is due to interference between
direct ($Z_2$ exchange) and indirect effects. 
These effects interfere constructively at some values of $\sin\xi$,
and destructively at others.

In our input scheme, for a fixed non-zero value of $\xi$, 
the gauge coupling $g_Z$ increases with $M_2$. 
This is due to the increase of $\rho_M$, and hence of $\rho$ 
[cf.\ Eqs.~(\ref{Eq.:gZ}), (\ref{rhom}) and (\ref{Del-rho})] 
with $M_2$.
This leads to a deviation of the cross section away from the SM value.
Thus, at large values of $M_2$, the $e^+e^-\to f\bar f$ cross section
is seen to impose strong constraints on the allowed mixing angle;
there is a narrowing, at large $M_2$, of the allowed region in $\sin\xi$.
(If we had frozen $\rho$ and $g_Z$ in our calculations,
then the different points in these figures would correspond
to different choices of Higgs sector---which would be rather unnatural.)

For the case of $b\bar b$ final states, we consider two different 
centre-of-mass energies, $\sqrt{s}=$ 190 and 500~GeV, in 
Figs.~6 and 7, respectively. 
The sensitivity of LEP2 and the NLC to $\xi$ and $M_2$ depends on the
chiral property of the $Z'$ coupling. For vector-, axial vector- 
and left-handed couplings, LEP2 (and the NLC) will have more
sensitivity than LEP1 at negative values of $\xi$.
For right-handed couplings, the situation is reversed.
Thus, LEP2 and the NLC have the potential to observe effects
of such a $Z'$ in this channel with masses up to the order of 1~TeV.
Concerning $\tau^+\tau^-$ final states, it appears that studies at
higher energies cannot improve on the results obtained at LEP1,
see Fig.~8.

In Fig.~9 we consider $t\bar t$ final states at $\sqrt{s}=$ 500~GeV.
As compared with $b\bar b$ final states, the sensitivity is lower.
However, the interference effects are different, and it is therefore
complementary to the $b\bar b$ channel.

Comparing the contour levels for the different final states, one can
see that the tightest restrictions are obtained from the $b\bar b$
final state, while the $\tau^+\tau^-$ case is the least restrictive.
This difference can be understood from an analysis of the expression
(\ref{Eq:F1}). In comparing with the SM cross section, the important
difference between the final-state fermions is the electric charge,
which dominates the main SM contribution. Thus, for a given value of
($\xi,M_2$) away from the resonance $\sqrt{s}=M_2$, the
relative deviation from the SM will be largest for $b$ quarks 
(because of the small electric charge) and smallest for $\tau$ leptons
(because of the large electric charge).

\subsection{Four-fermion final states}
We now turn our attention to the process
$e^+e^-\to b\bar b\nu_e\bar\nu_e$, which is sensitive to the
three-boson couplings.
In the SM, there are in the unitary gauge 23 Feynman diagrams
that contribute to this process.
In the extended model, the number of diagrams is 48 (in the unitary
gauge), even if all the scalars are excluded.
However, deviations from the standard model occur only
for the case of a non-zero $ZZ'$ mixing\footnote{There is no 
tree-level $WWZ'$ coupling since we assume the $Z'$ is an $SU(2)$
singlet.
One should keep in mind that the one-loop Feynman diagrams
give rise to a $WWZ'$ vertex even if at the tree level the mixing
angle vanishes.}.
As a result of this mixing there appear three-boson couplings involving
the $Z_2$.

Experimentally this process can be investigated by the detection
of the $b\bar b$ pair with missing energy.
In this case one measures the cross section of the processes
$e^+e^-\to b\bar b\nu_i\bar\nu_i$, integrated over a suitable
kinematical region (with the neutrinos of all three
families in the final state).
However, following the work of Ref.~\cite{Kal}, we believe that 
the main contribution comes from the process with the 
$\nu_e\bar\nu_e$ pair in the final state.

The process has been studied using CompHEP,
which generates the Feynman diagrams (we omit virtual Higgs particles)
and evaluates the cross section.
This has been integrated over phase space
according to the cuts of Ref.~\cite{Kal}.
For the $b$ jets to be detectable, we require them to have
sufficient energy, to be away from the beam pipe, not too close
to each other, and not have an invariant mass close to the $Z_1$.
Furthermore, the missing momentum should have a large transverse
component and a low rapidity, and the undetected neutrinos should
not originate from a $Z_1$:
\begin{align}
\label{Eq:cuts}
&E_b>20~\mbox{GeV}, & &20^\circ<\theta_b<160^\circ, \nonumber \\
&\theta_{b\bar b}>20^\circ, & &|m_{b\bar b}-m_{Z_1}|<3\Gamma_Z 
\nonumber \\
&\pslash_{\rm T}>40~\mbox{GeV}, & &\eta(\pslash)<1, \nonumber \\
&|\mslash-m_{Z_1}|>5\Gamma_Z.
\end{align}

At a given energy, the cross section increases significantly with
increasing values of $M_2$, and also with $|\sin\xi|$, as is
shown in Fig.~10. 
The increase in the cross section seen at increasing values
of $M_2$ is basically due to the fact that the coupling
$g_Z$ increases through the increase of the $\rho_{\rm M}$ parameter,
as was discussed in Sect.~4.1.

For the cases of $\sqrt{s}=190$ and 500~GeV, and for
vector couplings, the modifications of the cross sections,
with respect to the Standard Model,
are given in Figs.~11 and 12, respectively.
The gross features of these figures are rather similar to
those for the $b\bar b$ final states, and the sensitivity is quite
comparable.

\section{Concluding remarks}
\setcounter{equation}{0}
We have shown that a $Z'$ boson coupled only to the third-family 
fermions is rather difficult to discover,
even at high-energy colliders, thus confirming the more exploratory
analysis of Ref.~\cite{FWW}.

For hadronic colliders, such as the Tevatron or the LHC,
this $Z'$ is invisible because of the QCD background, which is 
many orders of magnitude greater than the cross sections
involving the $Z'$.
The data available from LEP1 already exclude significant regions
of the parameter space ($\xi$, $M_2$).
LEP2 and future $e^+e^-$ linear colliders can improve on these bounds,
in particular from studies of final states involving $b\bar b$.

It seems that some additional progress may be achieved
in the study of processes with four fermions in the final state,
if one investigates not only the full cross sections,
but also their dependence from
the  $b\bar{b}$ ($t\bar{t}$, $\tau\bar{\tau}$) invariant mass.
We hope to return to this question in future work.

\medskip
\section*{Acknowledgements}
It is a pleasure to thank Alexander Pukhov for instructing us on
how to use CompHEP, and Wolfgang Hollik for valuable discussions.
This research has been supported by the Research Council of Norway,
and by the Nordic project (NORDITA) `Fundamental constituents of matter'.
Also, A.A.\ and N.R.\ are partially supported by grant RFFI 98-02-18137.
\newpage
\appendix
\section*{Appendix~A:
Higgs sector and anomalies}
\renewcommand{\thesection}{A}
\setcounter{equation}{0}
Since we consider our model as an extension of the Standard Model, it
should contain the standard Higgs doublet:
\begin{equation}
\label{Eq:first}
H=\frac{1}{\sqrt{2}} 
\left(\begin{matrix}h^+ \\ v + h^0 \end{matrix}\right)
\end{equation}
where $v$ is the vacuum expectation value.
This field realizes the representation (1/2,1,0)
of the gauge group
$SU(2)_\Left\times U(1)_Y\times U(1)_{Y'}$
and hence has the following covariant derivative:
\begin{equation}
D_{\mu}H
=\left(\partial_{\mu}+igT^a W_{\mu}^a +i\frac{g'}{2}B_{\mu}\right)H.
\end{equation}

The kinetic term for this field contributes to the mass
terms of the $W^{\pm}$ and $Z$ bosons:
\begin{eqnarray}
|D_{\mu}H|^2 &\Longrightarrow& \frac{v^2}{8}(gW^3_{\mu}-g'B_{\mu})^2
+\frac{v^2}{2}g^2 W_{\mu}^+ W^{\mu}{}^- \nonumber \\
&=&
\frac{v^2}{8}\frac{g^2}{\cos^2\thW} Z_{\mu}Z^\mu
+\frac{v^2}{2}g^2 W_{\mu}^+ W^{\mu}{}^- \, .
\end{eqnarray}
This field does not supply the additional gauge boson $Z'$ with a mass.
It is  impossible to assign to the field $H$ a non-zero $Y'$ hypercharge: 
in this case the mass terms for
the fermions of the first two families will lose the
$U(1)_{Y'}$ invariance. Therefore $Y'_H=0$ and in our model
the field $H$ could give rise to the masses of the third-family
fermions only in the case when $Y'_{f\Left}=Y'_{f\Right}$ 
(not necessarily zero).
In the general case, when $Y'_{f\Left}\ne Y'_{f\Right}$, the ordinary Higgs
field $H$ does not contribute to the masses of the third-family
fermions.

The simplest way to obtain a $Z'$ mass is to introduce a scalar singlet
$\phi$, which transforms as $(0,0,Y'_S)$ and has a covariant derivative:
\begin{equation}
D_{\mu}\phi=\left(\partial_{\mu}+ig_{Z'} \frac{Y_S'}{2}Z_{\mu}'\right) \phi.
\end{equation}
With a non-vanishing vacuum expectation value,
$\langle\phi\rangle_{\rm vac}=v_S/\sqrt{2}$,
it produces the following mass term for the $Z'$ boson:
\begin{equation}
|D_{\mu}\phi|^2 \Longrightarrow \frac{g_{Z'}^2}{8}Y_S'{}^2v_S^2
Z'_{\mu}{}^2.
\end{equation}

By means of this field $\phi$, an arbitrary mass can produced
for the $Z'$ boson, but it is not possible to get
$ZZ'$  mixing at the tree level. 
In the case where the $Z'$ boson is not a pure vector particle
[$Y'_{f\Left}\ne Y'_{f\Right}$], it is also impossible to obtain
gauge-invariant mass terms for the fermions of the third family
(see above). 

In order to solve these two problems, let us introduce an additional
Higgs doublet $H_1$, which interacts with the $Z'$ and hence has a
non-zero $Y'$ hypercharge. This field transforms as 
$(\frac{1}{2},1,Y'_1)$ and has a covariant derivative:
\begin{equation}
D_{\mu}H_1 =\left(\partial_{\mu}+igT^aW_{\mu}^a
+i\frac{g'}{2}B_{\mu}+ i\frac{g_{Z'}}{2}Y'_1 Z'_{\mu}\right)H_1.
\end{equation}
(The value of $Y'_1$ is not arbitrary when the field
$H_1$ gives rise to fermion masses, this will be
discussed at the end of this appendix.)
With $\langle H_1\rangle_{\rm vac}=v_1/\sqrt{2}$,
it produces the following contribution to the mass term of the neutral
gauge bosons:
\begin{equation}
|D_{\mu}H_1|^2 \Longrightarrow \frac{g^2}{\cos^2\thW}
\frac{v_1^2}{8}\left(Z^2_{\mu}+\frac{g_{Z'}^2}{g'{}^2} \sin^2\thW
Y_1'{}^2 Z_{\mu}'{}^2
- 2\,\frac{g_{Z'}}{g'} \sin\thW Y'_1 Z_{\mu} Z'{}^{\mu}\right).
\end{equation}
So in this case when all three Higgs fields are present, one obtains
a mass matrix for the neutral gauge bosons in the general form:
\begin{equation}
(Z,Z') \left( \begin{matrix}M^2_Z   & M^2_{ZZ'} \\
                          M^2_{ZZ'} & M^2_{Z'}
              \end{matrix} \right)
\left(\begin{matrix}Z \\ Z'\end{matrix}\right),
\end{equation}
where
\begin{equation}
M^2_Z=\frac{g^2}{4\cos^2\thW}(v^2 +v_1^2); \qquad
M^2_{Z'}=\frac{g_{Z'}^2}{4}(v_S^2Y_S^2 +v_1^2Y_1^2), 
\end{equation}
\begin{equation}
M^2_{ZZ'}=   -\frac{g^2}{\cos^2\thW}\frac{v^2}{4}
\frac{g_{Z'}}{g'}\sin\thW Y_1.
\end{equation}

After the diagonalization of this matrix, the mass eigenstates 
$M_1$, $M_2$ and the mixing angle $\xi$ can be expressed through
v.e.v.'s of the introduced scalars.
The field $H_1$ also gives masses to the third-family fermions.

However it should be stressed that the values of the $Y'$ hypercharge
are not arbitrary: restrictions come from the triangle anomalies.
The well-known condition of ABJ anomalies reads \cite{DelSal,AlvWit}:
\begin{equation}
\Tr(T_i \{ T_j,T_k\})_\Left=\Tr(T_i \{ T_j,T_k\})_\Right
\end{equation}
where $T$ are the matrices of the fermion representations, and
$i$, $j$, $k$ may refer to different subgroups of the full gauge group.

In the Standard Model with the group
$SU(3)_c\times SU(2)_\Left\times U(1)_Y$,
these conditions unambiguously fix
ratios of electric charges (or hypercharges) of fermions:
four independent conditions of diagrams with
LL$Y$, $YYY$, $ccY$ and $ggY$ external fields (here L denotes
the $SU(2)_\Left$ field, $Y$ the $U(1)_Y$ field,
$c$ the $SU(3)$ field, and $g$ the graviton) for
four ratios of $Y_{l\Left}$, $Y_{e\Right}$,
$Y_{q\Left}$, $Y_{u\Right}$, $Y_{d\Right}$
have a single solution. 
For the extended model, there are additional
conditions: LL$Y'$, $YYY'$, $YY'Y'$, $Y'Y'Y'$, $ccY'$ and $ggY'$, which 
necessarily lead to $Y'=Y$.
All other $Y'$ assignments need an extension of the fermionic
sector. In order not to introduce exotic fermions let us
consider the simplest extension of the fermionic sector, namely
addition of right-handed neutrinos.
In this case the anomaly cancellation conditions, involving
the $Z'$ boson read:
\begin{eqnarray}
\Left\Left Y': 
      &\quad & Y'_{\calL\Left}+3Y'_{\calQ\Left}=0, \nonumber \\[2mm]
ccY': &\quad & 
3(2Y'_{\calQ\Left}-Y'_{t\Right}-Y'_{b\Right})=0, \nonumber \\[2mm]
ggY': &\quad & 2Y'_{\calL\Left}-Y'_{\tau\Right}-Y'_{\nu\Right}
     +3(2Y'_{\calQ\Left}-Y'_{t\Right}-Y'_{b\Right})=0, \nonumber \\[2mm]
Y'Y'Y': &\quad & 2Y'_{\calL\Left}{}^3-Y'_{\tau\Right}{}^3
     -Y'_{\nu \Right}{}^3
     +3(2Y'_{\calQ\Left}{}^3-Y'_{t\Right}{}^3-Y'_{b\Right}{}^3)=0,
     \nonumber \\[2mm]
YYY': &\quad &  2Y'_{\calL\Left}-4Y'_{\tau\Right}
     +3\left(
     \frac{2}{9}Y'_{\calQ\Left}-\frac{16}{9}Y'_{t\Right}
     -\frac{4}{9}Y'_{b\Right} \right)=0, \nonumber \\[2mm]
YY'Y': &\quad &  -2Y'_{\calL\Left}{}^2+2Y'_{\tau\Right}{}^2
     +3\left(\frac{2}{3}Y'_{\calQ\Left}{}^2 -\frac{4}{3}Y'_{t\Right}{}^2
     +\frac{2}{3}Y'_{b\Right}{}^2 \right)=0. \nonumber \\
\end{eqnarray}
Here, $\calL$ and $\calQ$ denote the third-family 
lepton and quark doublets.
The last two equations are not independent.
With $x_i \equiv Y'_i/Y'_{\calL\Left}$,
the system has the following solution: 
\begin{eqnarray}
\label{Eq:AppAvalues}
& & x_{\calL\Left}=1, \qquad
x_{\calQ\Left}= -\frac13, \qquad
x_{\tau\Right}=2-x_{\nu \Right},  \nonumber \\
& & x_{b\Right}=\frac23-x_{\nu \Right}, \qquad
x_{t\Right}=-\frac43+x_{\nu\Right},
\end{eqnarray}
where $Y'_{\nu\Right}$ (or $x_{\nu\Right}$) remains undetermined.
For the case $x_{\nu\Right}=0$ one gets $Y=Y'$ (case of the SM), 
whereas $x_{\nu\Right}=1$ corresponds to a purely vector $Z'$ boson;
the general case can be treated as a linear combination of 
these two cases.

These values of fermionic hypercharges $Y'_i$ determine
the hypercharge $Y'_1$ of the $H_1$ field: the gauge invariance
of the fermionic mass term necessarily leads to
$Y'_1=Y'_{\tau \Left}-Y'_{\tau \Right}$.

A vector-like $Z'$-fermion coupling is achieved with a universal choice:
$x_{\nu \Right}=1$.
However, as follows from Eqs.~(\ref{Eq:AppAvalues}), 
the assignment for the $Z'$-fermion vertex to be
axial, left- or right-handed cannot be done universally. 
For instance, the coupling may be purely axial either for
$b$ quarks, or for $\tau$ leptons, or for $t$ quarks, 
but not simultaneously.

Let us now comment on the chirality choices considered in Figs.~6--9.
For the $Z'b\bar b$ coupling (Figs.~6 and 7) to be vector, axial,
left- or right-handed, one must choose $x_{\nu \Right}=1$, $1/3$,
$2/3$, or $-\infty$, respectively.
For the $Z'\tau^+\tau^-$ coupling (Fig.~8) to be vector, axial,
left- or right-handed, one must choose $x_{\nu \Right}=1$, 
$3$, $2$,  or $-\infty$, respectively.
For the $Z't\bar t$ coupling (Fig.~9) to be vector, axial,
left- or right-handed, one must choose $x_{\nu \Right}=1$, 
$5/3$, $4/3$, or $\infty$, respectively.

It may be useful to notice that the cancellation of anomalies
is necessary only if the additional $Z'$ boson is treated as
a fundamental particle. For a composite $Z'$, arbitrary values
of fermionic hypercharge are allowed.

\section*{Appendix~B: QCD background}
\renewcommand{\thesection}{B}
\setcounter{equation}{0}
We shall assume that light-flavour quark jets can be rejected
and do not constitute a background. Thus, the background of interest
is the one due to $b\bar b$ (and, at sufficiently high energies, 
$t\bar t$) jets.
\subsection{Quark--antiquark annihilation}
With $p_1+p_2=b+\bar b$, $\hat s=(p_1+p_2)^2$, 
the matrix element can be written as
\begin{equation}
M=\bar v(p_2)\gamma^\mu gT_a u(p_1)
\frac{-g_{\mu\nu}}{\hat s}
\bar u(b)\gamma^\nu gT_b v(\bar b)\, \delta_{ab},
\end{equation}
with $g$ the QCD coupling.
Properly averaged over spin and colour,
the squared matrix element takes the form:
\begin{equation}
\overline{|M|^2}
=\frac{8}{9}\, \frac{g^4}{\hat s^2}
\{2(\hat t -m_i^2-m_b^2)^2 +2(\hat u -m_i^2-m_b^2)^2
+(m_i^2+m_b^2)\hat s\},
\end{equation}
where $\hat t=(p_1-b)^2$, $\hat u=(p_1-\bar b)^2$, 
and $p_1^2=p_2^2=m_i^2$.
\subsection{Gluon fusion}
Gluon fusion will also lead to $b\bar b$ (or $t\bar t$) jets.
There are three diagrams contributing, including the one due
to the three-gluon vertex ($p_1+p_2=p=b+\bar b$):
\begin{eqnarray}
M  &=&M_1+M_2+M_3, \nonumber \\
M_1&=&\bar u(b)ig\epsslash_1 T_a\,
\frac{i(\bslash-\pslash_1+m_b)}{(b-p_1)^2-m_b^2}\,
ig\epsslash_2 T_b\, v(\bar b), \nonumber \\
M_2&=&\bar u(b)ig\epsslash_2 T_b\,
\frac{i(\bslash-\pslash_2+m_b)}{(b-p_2)^2-m_b^2}\,
ig\epsslash_1 T_a\, v(\bar b), \nonumber \\
M_3&=&\bar u(b)ig\gamma_\mu T_c v(\bar b)\,
\frac{g^{\mu\nu}-\xi p^\mu p^\nu/p^2}{p^2+\ieps} \nonumber \\
& & \times (-ig)f^{abc}
[g_{\alpha\beta}(p_1-p_2)_\nu +g_{\beta\nu}(p_2+p)_\alpha 
-g_{\nu\alpha}(p+p_1)_\beta]\epsilon_1^\alpha \epsilon_2^\beta.
\nonumber \\
\end{eqnarray}
Since the final-state quark and antiquark are taken to be on-shell,
the gauge-dependent part of $M_3$ vanishes.
Properly averaged over spin and colour,
the squared matrix element takes the form
\begin{equation}
\overline{|M|^2}
=\frac{g^4}{2}\bigl[X+Y+Z\bigr],
\end{equation}
where $X$ is due to $M_1$ and $M_2$, $Y$ is due to their interference
with $M_3$, and $Z$ represents $M_3$ squared. Furthermore,
\begin{equation}
X=\frac{1}{(\hat t-m_b^2)^2}\, X_{tt} 
+\frac{1}{(\hat t-m_b^2)(\hat u-m_b^2)}\, X_{tu}
+\frac{1}{(\hat u-m_b^2)^2}\, X_{uu},
\end{equation}
with
\begin{eqnarray}
X_{tt}&=&\frac{8}{3}\,\bigl[(\hat t+m_b^2)(\hat u-m_b^2)
+2m_b^2(\hat s+2m_b^2)\bigr],
\nonumber \\
X_{tu}&=&-\frac{2}{3}\, m_b^2(\hat s-4m_b^2), \nonumber \\
X_{uu}&=&\frac{8}{3}\,\bigl[(\hat t-m_b^2)(\hat u+m_b^2)
+2m_b^2(\hat s+2m_b^2)\bigr].
\end{eqnarray}
Similarly,
\begin{equation}
Y=Y_t+Y_u,
\end{equation}
with
\begin{eqnarray}
Y_t&=&\frac{-6}{\hat s(\hat t-m_b^2)}
\bigl[(\hat t-m_b^2)^2+\hat s m_b^2\bigr], \nonumber \\
Y_u&=&Y_t(\hat t\leftrightarrow \hat u)
\end{eqnarray}
and finally
\begin{equation}
Z=\frac{12}{\hat s^2}
\bigl[(\hat t-m_b^2)(\hat u-m_b^2)-\hat s(\hat s+2m_b^2)\bigr].
\end{equation}

\begin{figure}
\begin{center}
\setlength{\unitlength}{1cm}
\begin{picture}(14.3,16.0)
\put(0.,0.0){
\mbox{\epsfysize=14cm\epsffile{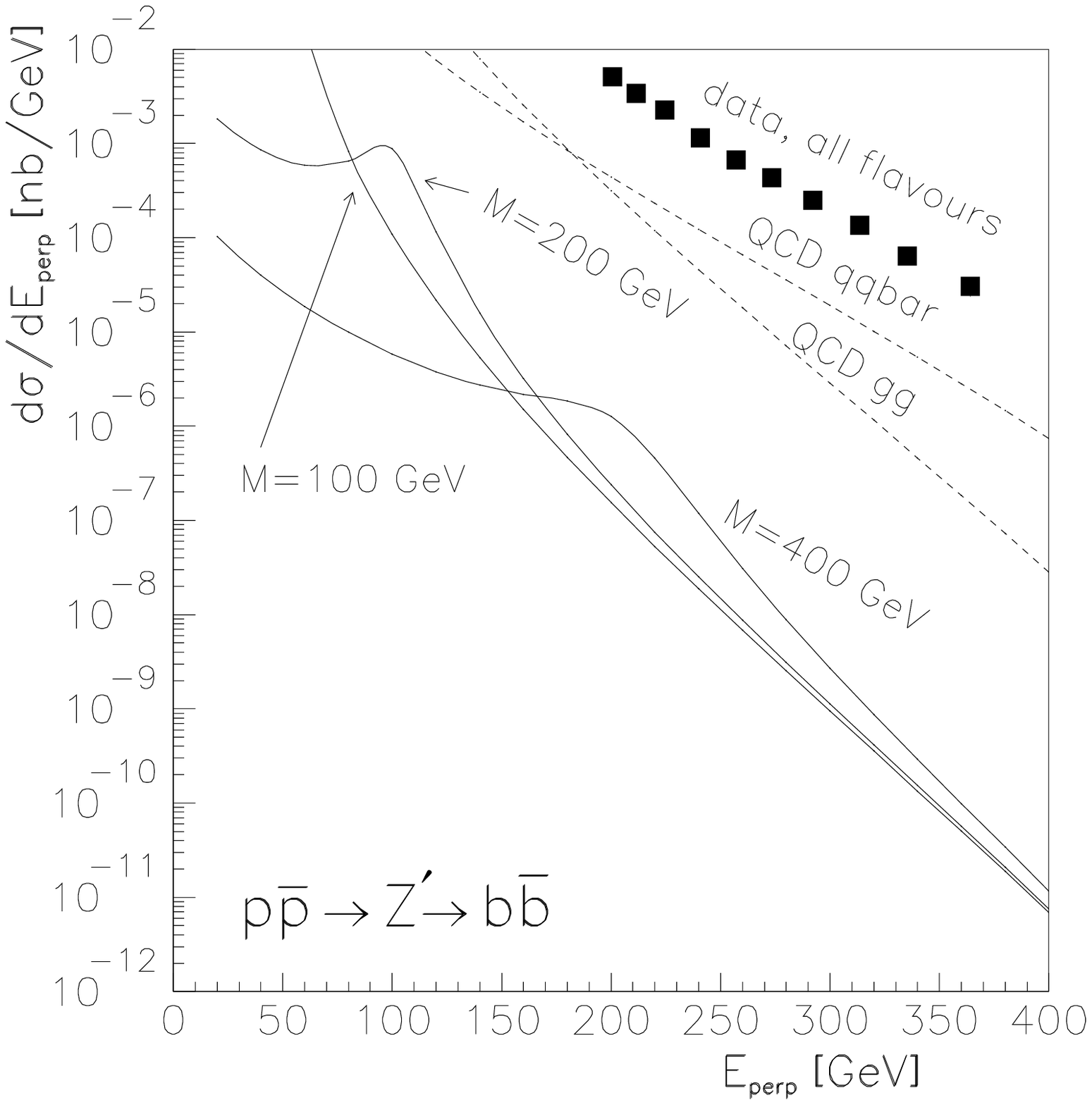}}}
\end{picture}
\begin{capt}
Cross sections for inclusive jet production at the Tevatron,
$p\bar p$ collisions at $E_{\rm cm}=1.8$~TeV.
The solid curves represent Drell--Yan-type production of
$Z'$, from $b$ and $\bar b$ (sea) quarks in the initial state.
Three masses are considered, $M_{Z'}=100$, 200 and 400~GeV.
The couplings are: $g_{Z'}v_b'=g_{Z'}a_b'=1$.
Also the contributions from the dominant QCD mechanisms are shown, 
as well as data (summed over all flavours) \cite{CDF}.
\end{capt}
\end{center}
\end{figure}

\begin{figure}
\begin{center}
\setlength{\unitlength}{1cm}
\begin{picture}(14.3,16.0)
\put(0.,0.0){
\mbox{\epsfysize=14cm\epsffile{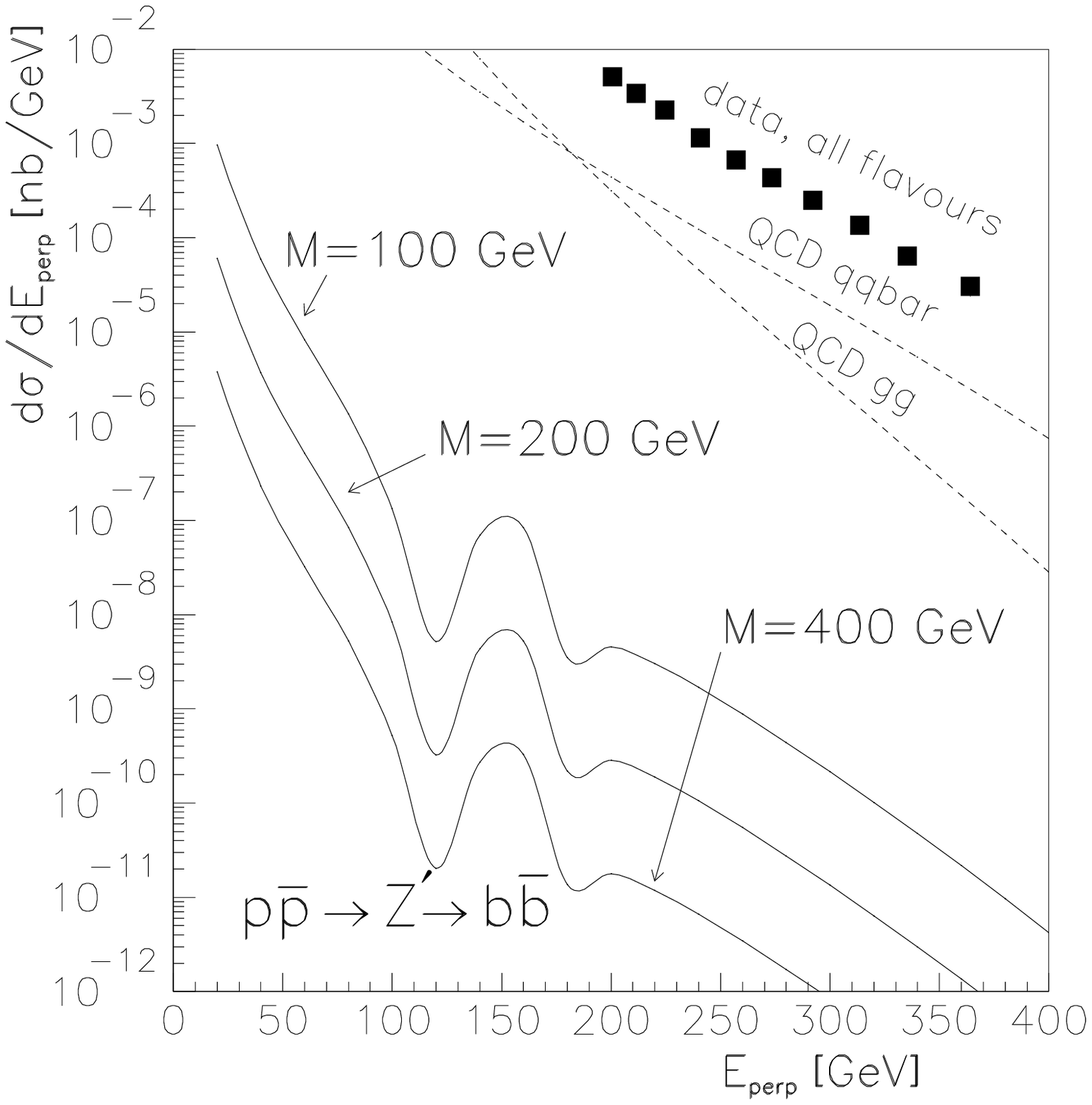}}}
\end{picture}
\begin{capt}
Cross sections for inclusive jet production at the Tevatron,
$p\bar p$ collisions at $E_{\rm cm}=1.8$~TeV.
The solid curves represent production of
$Z'$, from gluon fusion.
Three masses are considered, $M_{Z'}=100$, 200 and 400~GeV.
Both $b$ and $t$ quarks contribute to the triangle diagram,
with equal couplings, $g_{Z'}a_t'=g_{Z'}a_b'=1$.
Also the contributions from the dominant QCD mechanisms are shown, 
as well as data (summed over all flavours) \cite{CDF}.
\end{capt}
\end{center}
\end{figure}

\begin{figure}
\begin{center}
\setlength{\unitlength}{1cm}
\begin{picture}(14.3,16.0)
\put(0.,0.0){
\mbox{\epsfysize=14cm\epsffile{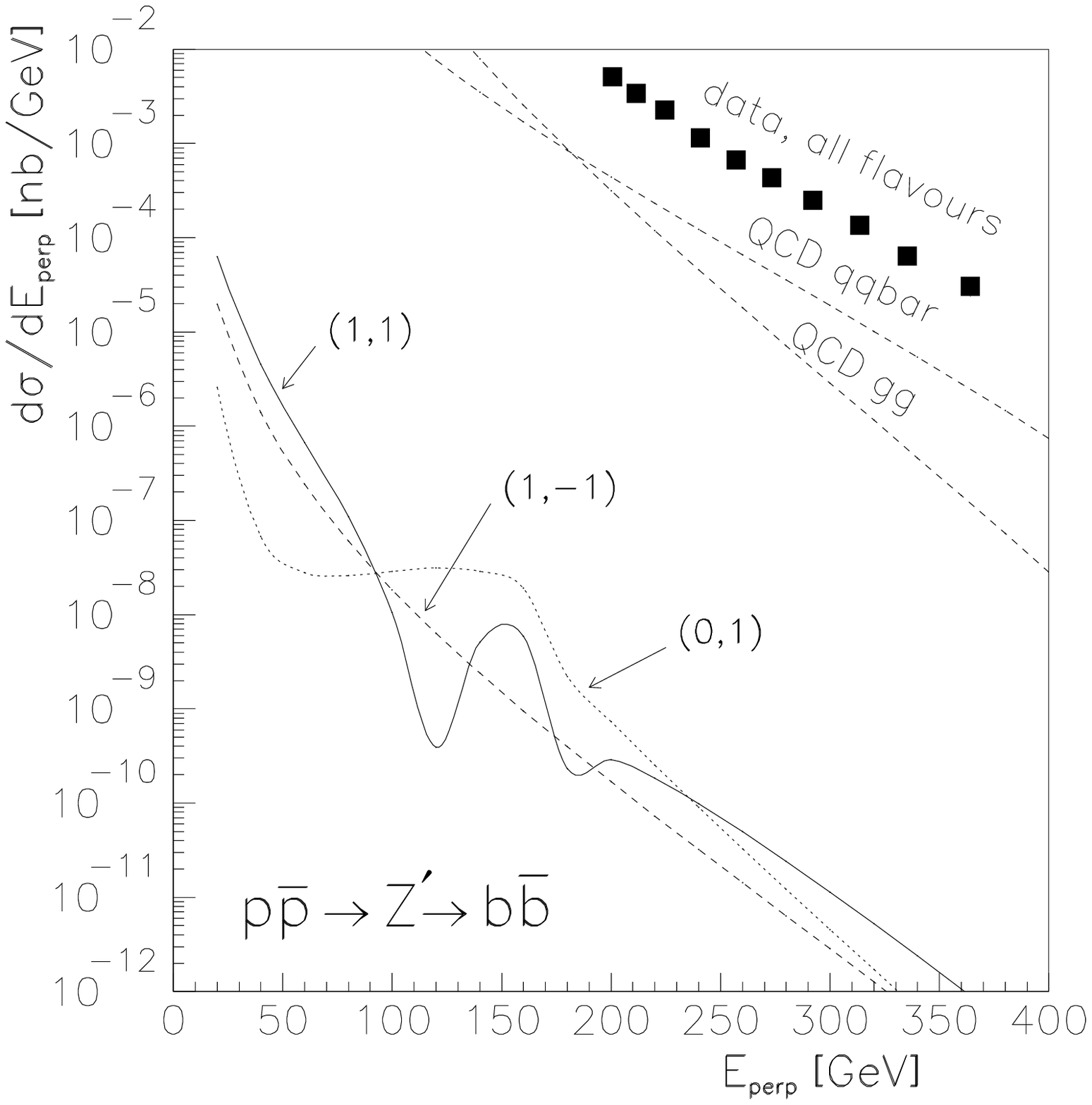}}}
\end{picture}
\begin{capt}
Similar to Fig.~1, but for $M_{Z'}=200$~GeV only.
Three cases of $q\bar qZ'$ couplings are considered:
($g_{Z'}a_t'$, $g_{Z'}a_b'$) $=$ ($1,1$) (solid), ($0,1$) (dotted),
and ($1,-1$) (dashed).
\end{capt}
\end{center}
\end{figure}


\begin{figure}
\begin{center}
\setlength{\unitlength}{1cm}
\begin{picture}(14.3,16.0)
\put(0.,0.0){
\mbox{\epsfysize=14cm\epsffile{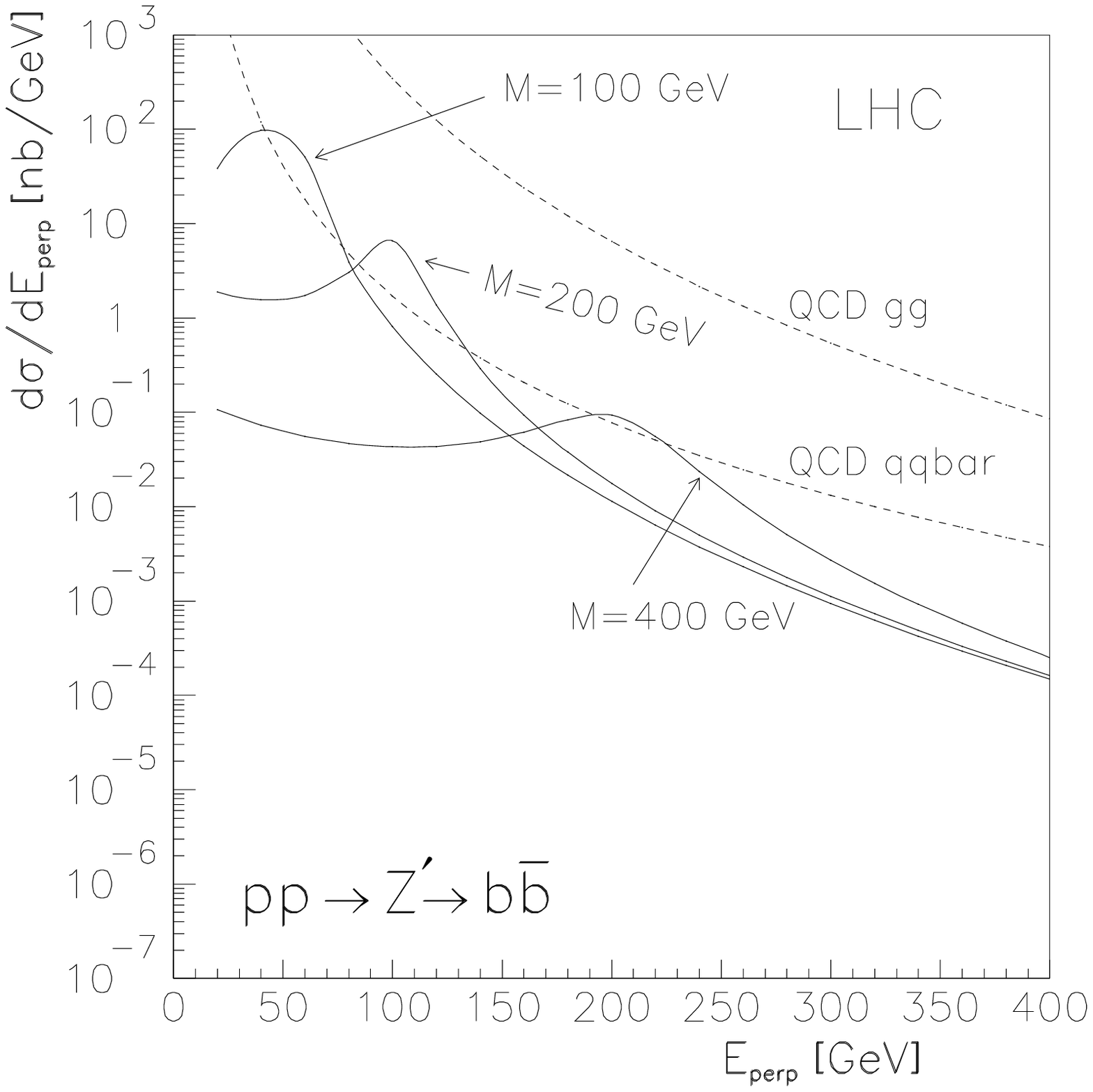}}}
\end{picture}
\begin{capt}
Cross sections for inclusive jet production at the LHC,
$pp$ collisions at $E_{\rm cm}=14$~TeV.
The solid curves represent Drell--Yan-type production of
$Z'$, from $b$ and $\bar b$ (sea) quarks in the initial state.
Three masses are considered, $M_{Z'}=100$, 200 and 400~GeV.
Also the contributions from the dominant QCD mechanisms are shown,
using standard distribution functions \cite{Plothow}.
\end{capt}
\end{center}
\end{figure}
\begin{figure}
\begin{center}
\setlength{\unitlength}{1cm}
\begin{picture}(14.3,16.0)
\put(0.,0.0){
\mbox{\epsfysize=14cm\epsffile{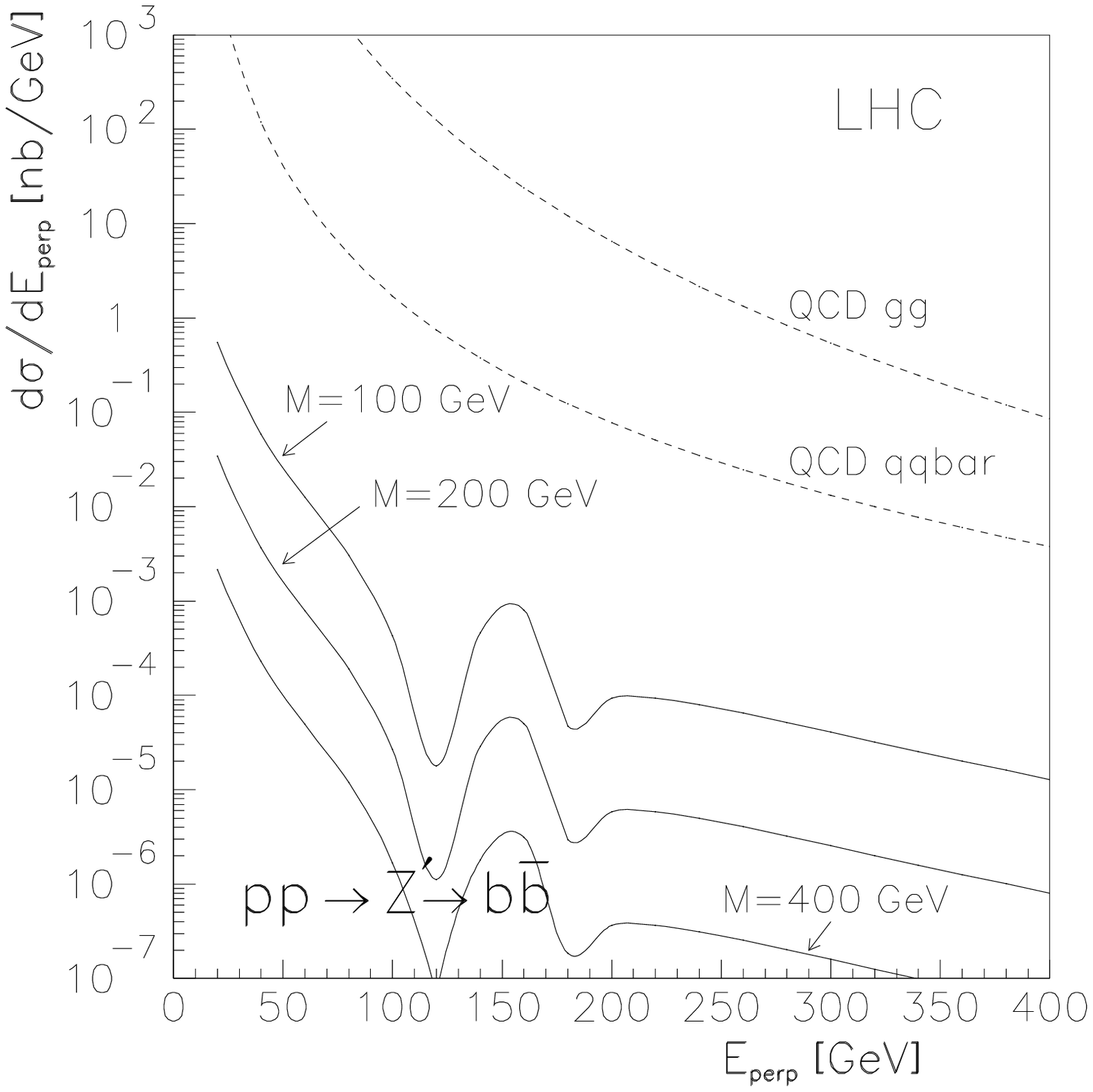}}}
\end{picture}
\begin{capt}
Cross sections for inclusive jet production at the LHC,
$pp$ collisions at $E_{\rm cm}=14$~TeV.
The solid curves represent production of
$Z'$, from gluon fusion.
Three masses are considered, $M_{Z'}=100$, 200 and 400~GeV.
Both $b$ and $t$ quarks contribute to the triangle diagram,
with equal couplings, $g_{Z'}a_t'=g_{Z'}a_b'=1$.
Also the contributions from the dominant QCD mechanisms are shown.
\end{capt}
\end{center}
\end{figure}

\begin{figure}
\begin{center}
\setlength{\unitlength}{1cm}
\begin{picture}(14.3,18.0)
\put(-1.,9.0){
\mbox{\epsfysize=8.5cm\epsffile{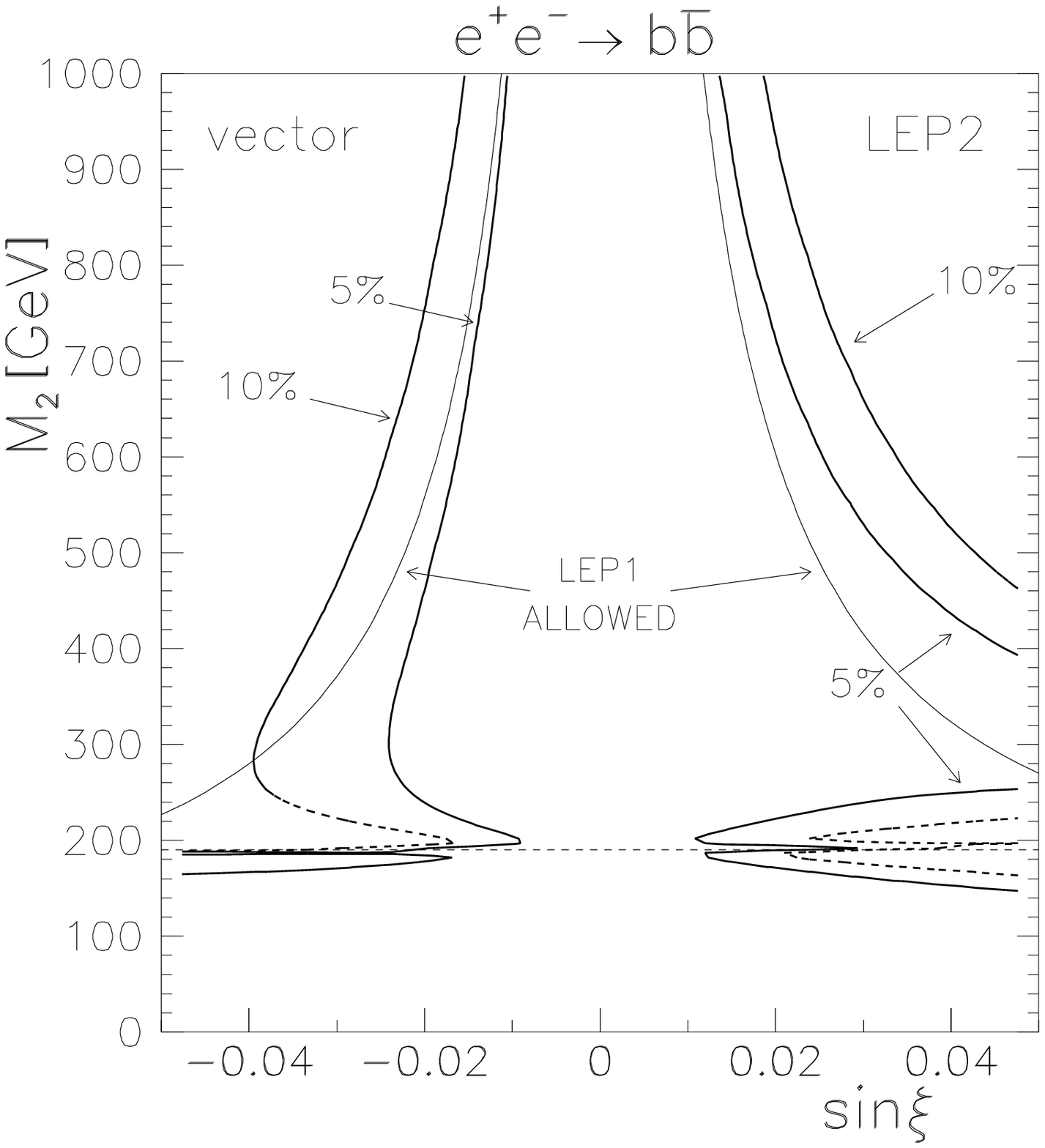}
      \epsfysize=8.5cm\epsffile{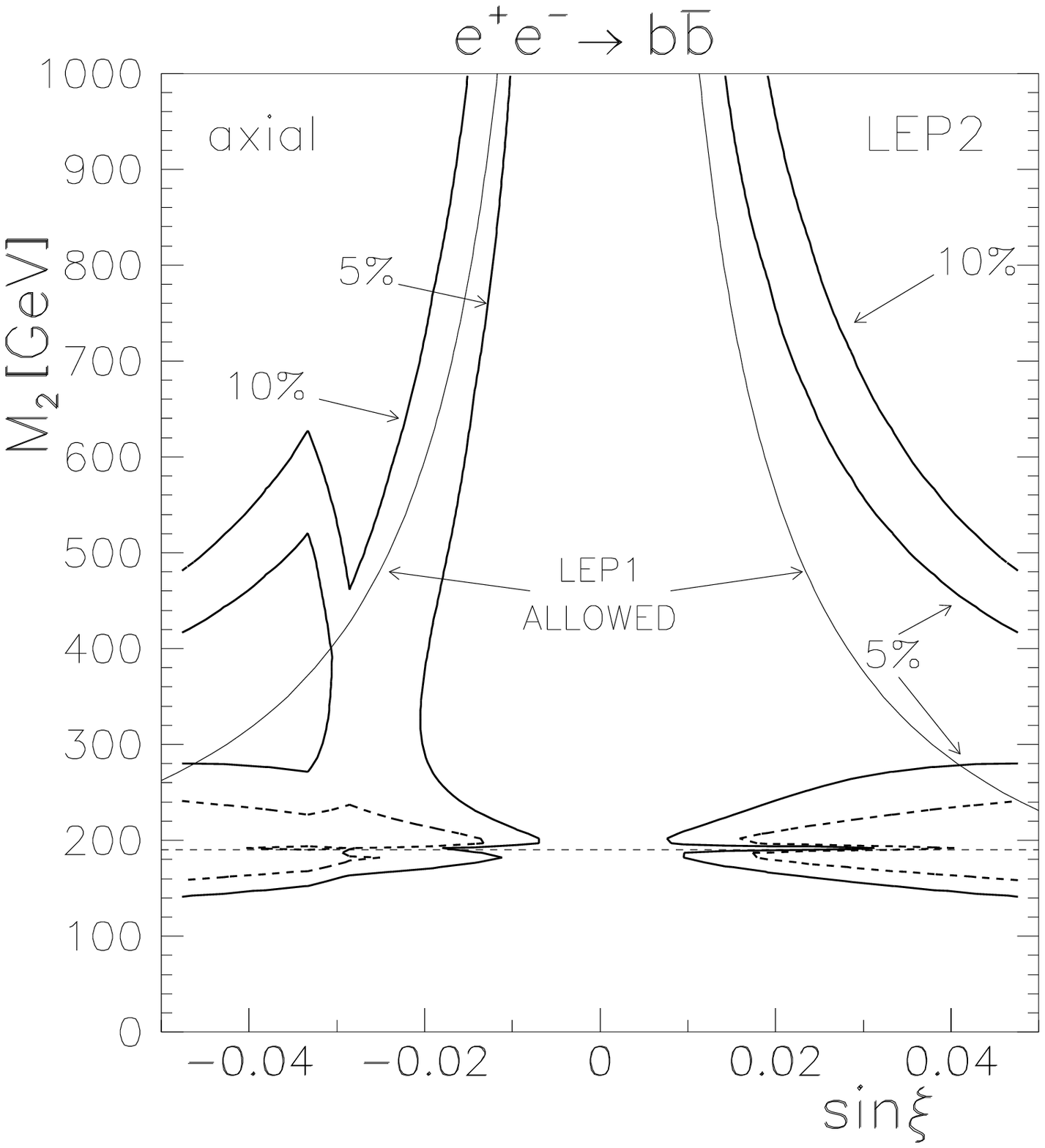}}}
\put(-1.,0.0){
\mbox{\epsfysize=8.5cm\epsffile{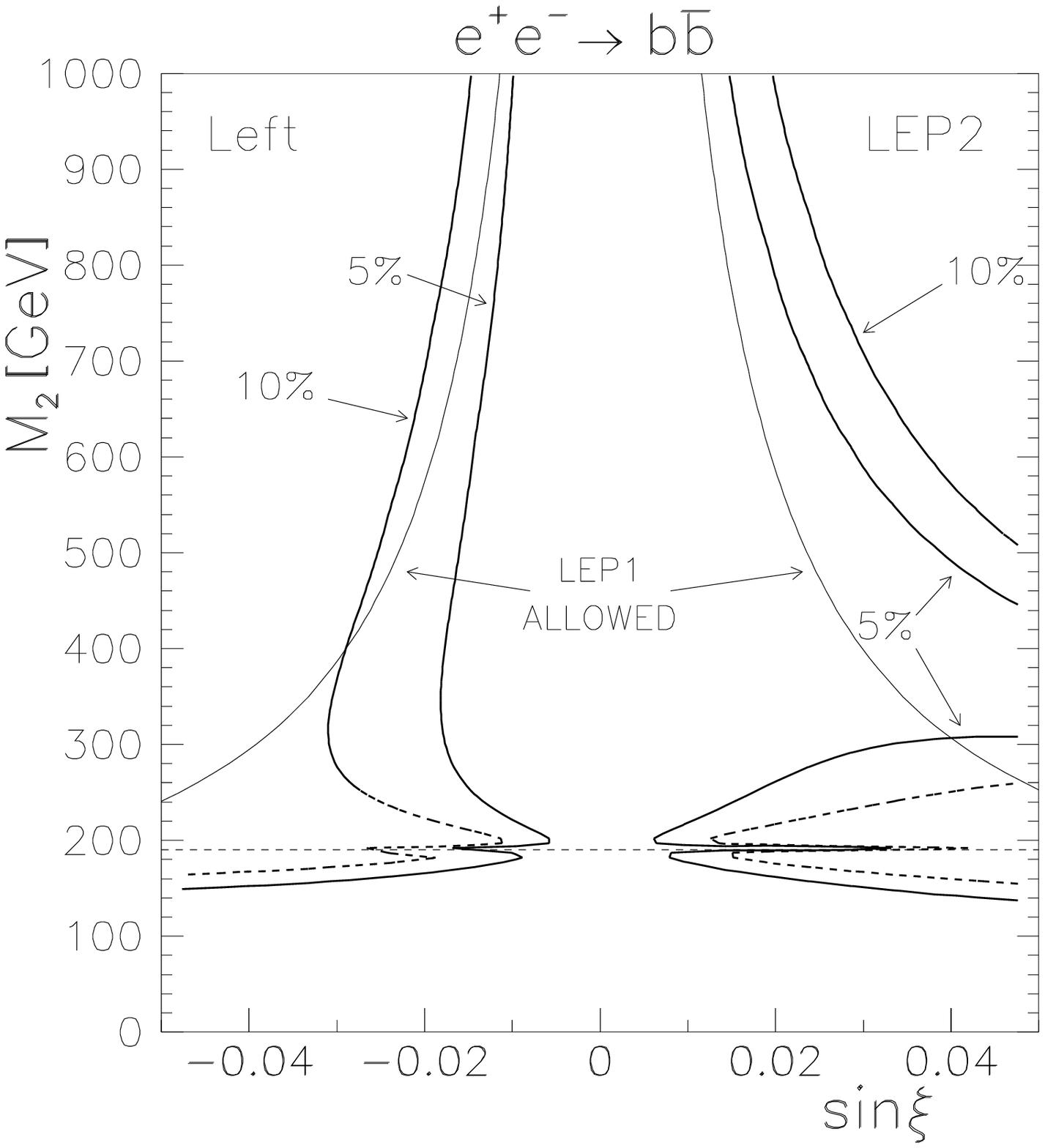}
      \epsfysize=8.5cm\epsffile{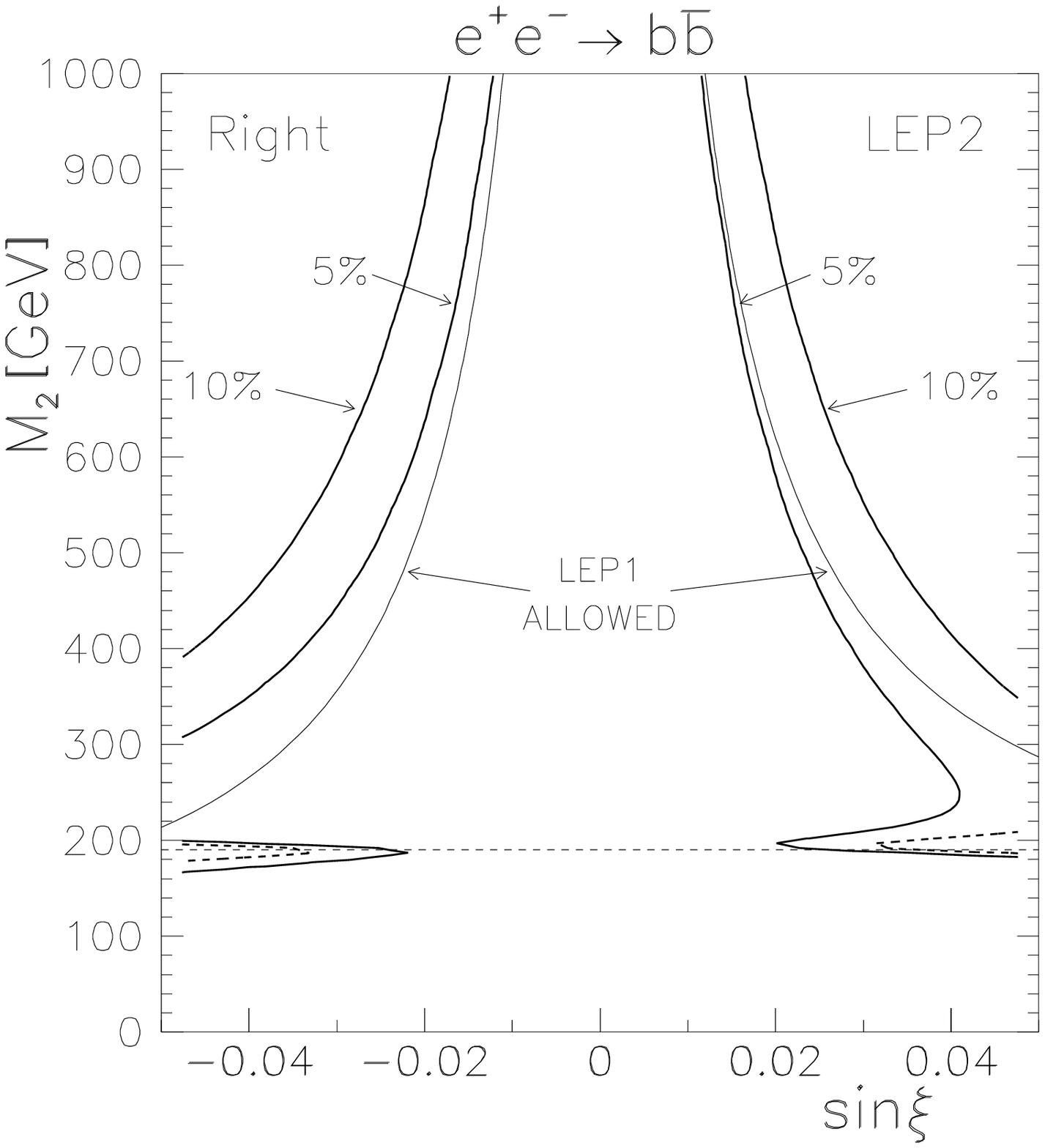}}}
\end{picture}
\begin{capt}
Allowed regions of $\sin\xi$ and $M_2$ obtained from LEP1 data
(95\% C.L.) for the process $e^+e^-\to b\bar b$.
Also shown are bounds anticipated from LEP2 at levels of assumed precision
as indicated by labels. 
Four different chiralities are considered.
\end{capt}
\end{center}
\end{figure}

\begin{figure}
\begin{center}
\setlength{\unitlength}{1cm}
\begin{picture}(14.3,18.0)
\put(-1.,9.0){
\mbox{\epsfysize=8.5cm\epsffile{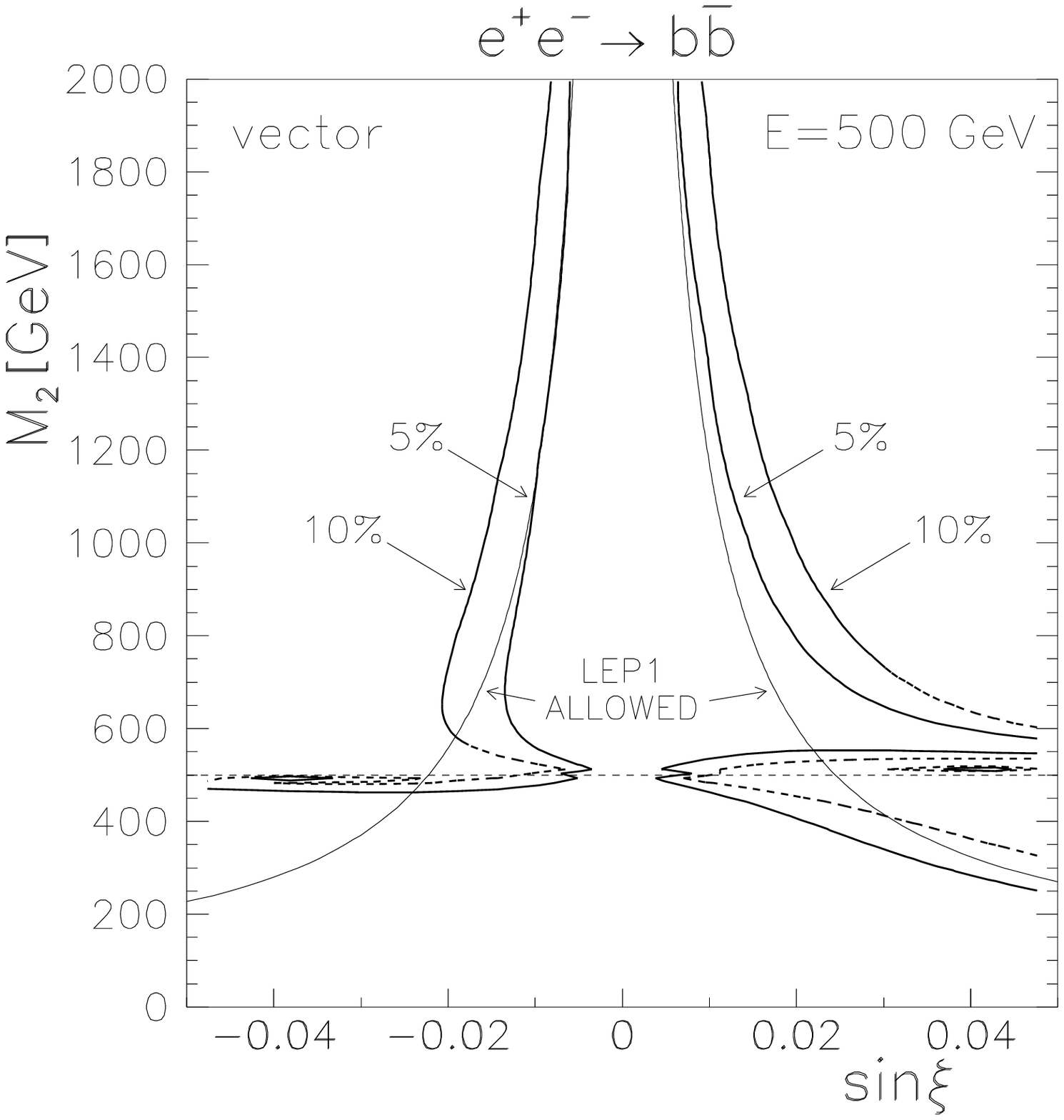}
      \epsfysize=8.5cm\epsffile{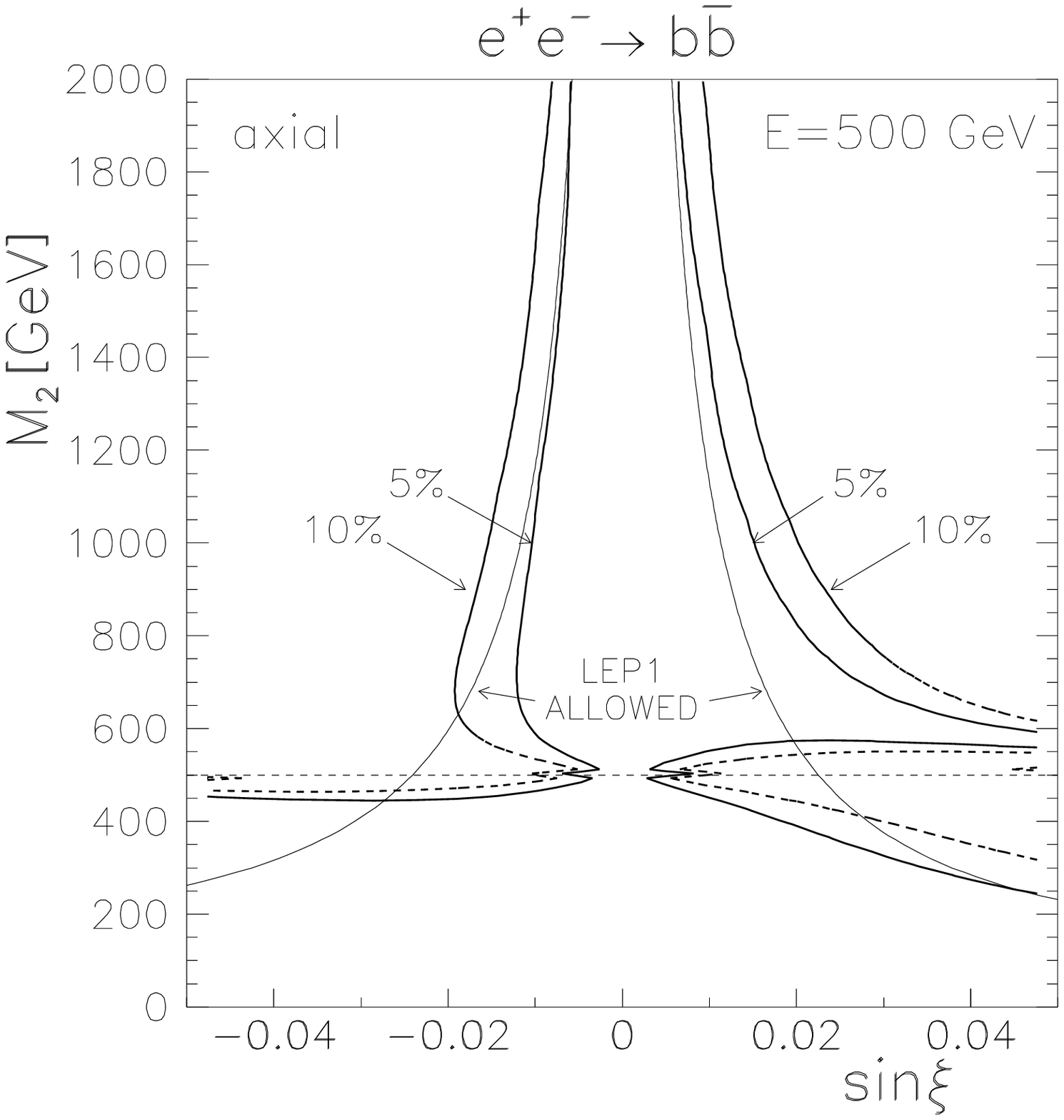}}}
\put(-1.,0.0){
\mbox{\epsfysize=8.5cm\epsffile{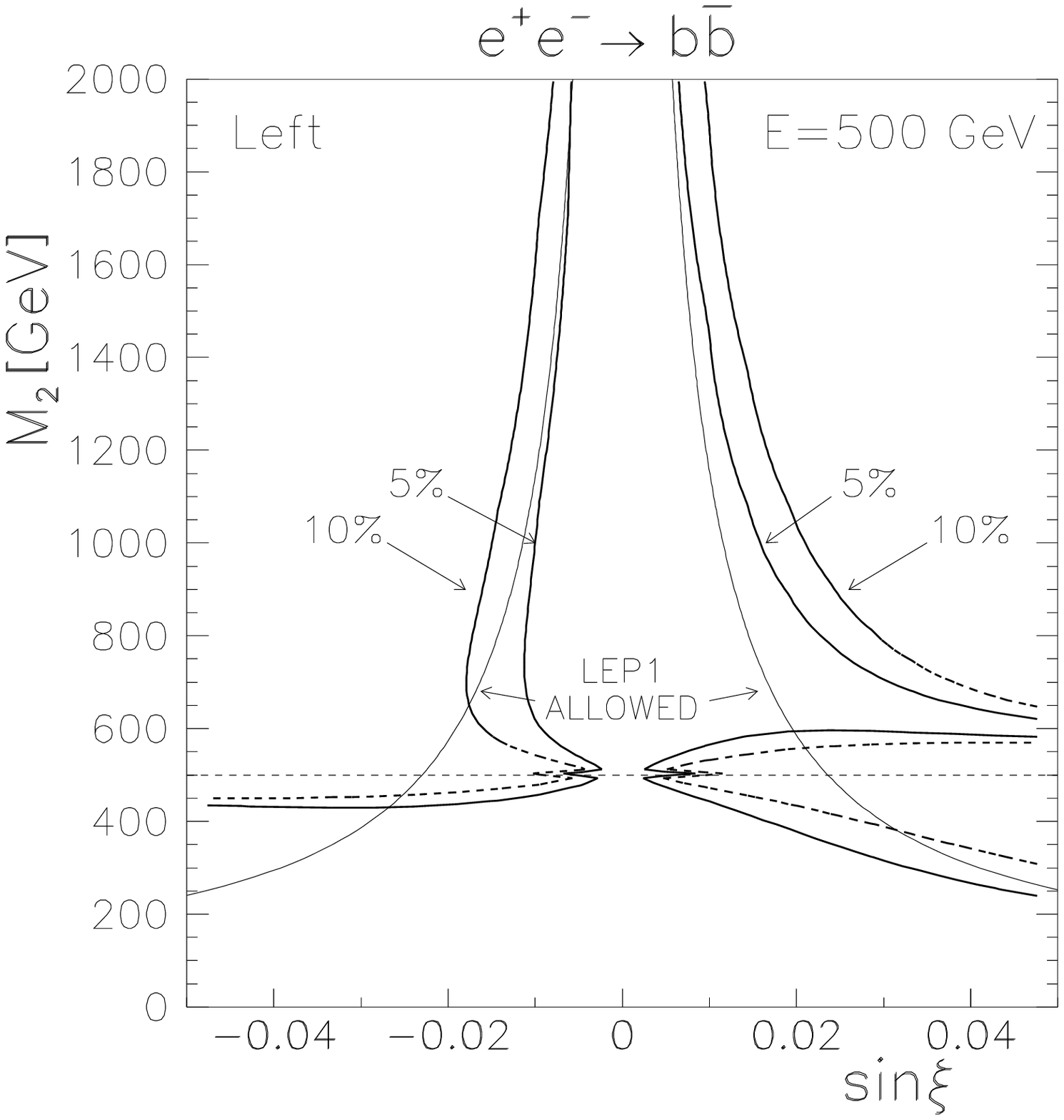}
      \epsfysize=8.5cm\epsffile{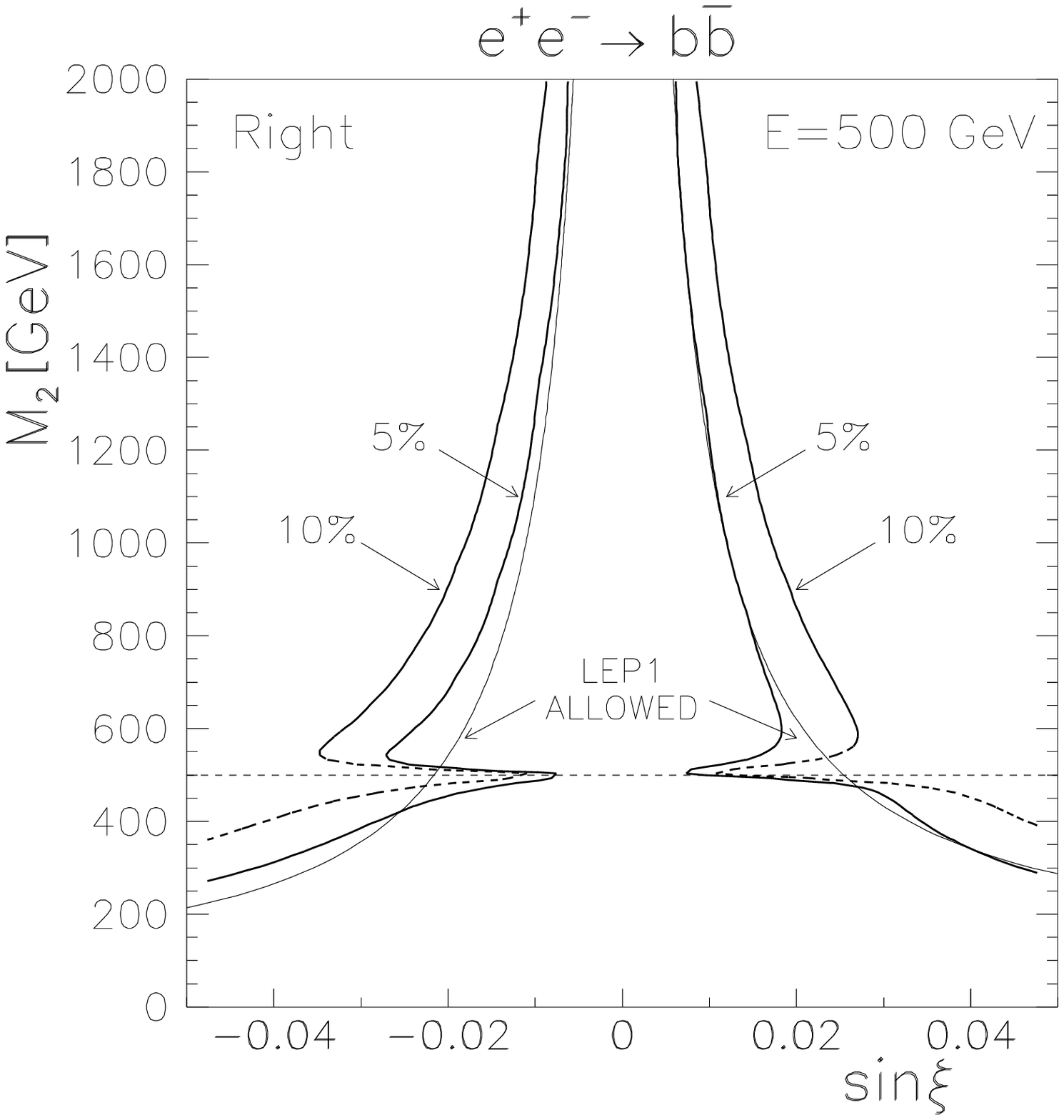}}}
\end{picture}
\begin{capt}
Allowed regions of $\sin\xi$ and $M_2$ obtained from LEP1 data
(95\% C.L.) for the process $e^+e^-\to b\bar b$.
Also shown are bounds anticipated from 500~GeV
at levels of assumed precision as indicated by labels. 
\end{capt}
\end{center}
\end{figure}

\begin{figure}
\begin{center}
\setlength{\unitlength}{1cm}
\begin{picture}(14.3,18.0)
\put(-1.,9.0){
\mbox{\epsfysize=8.5cm\epsffile{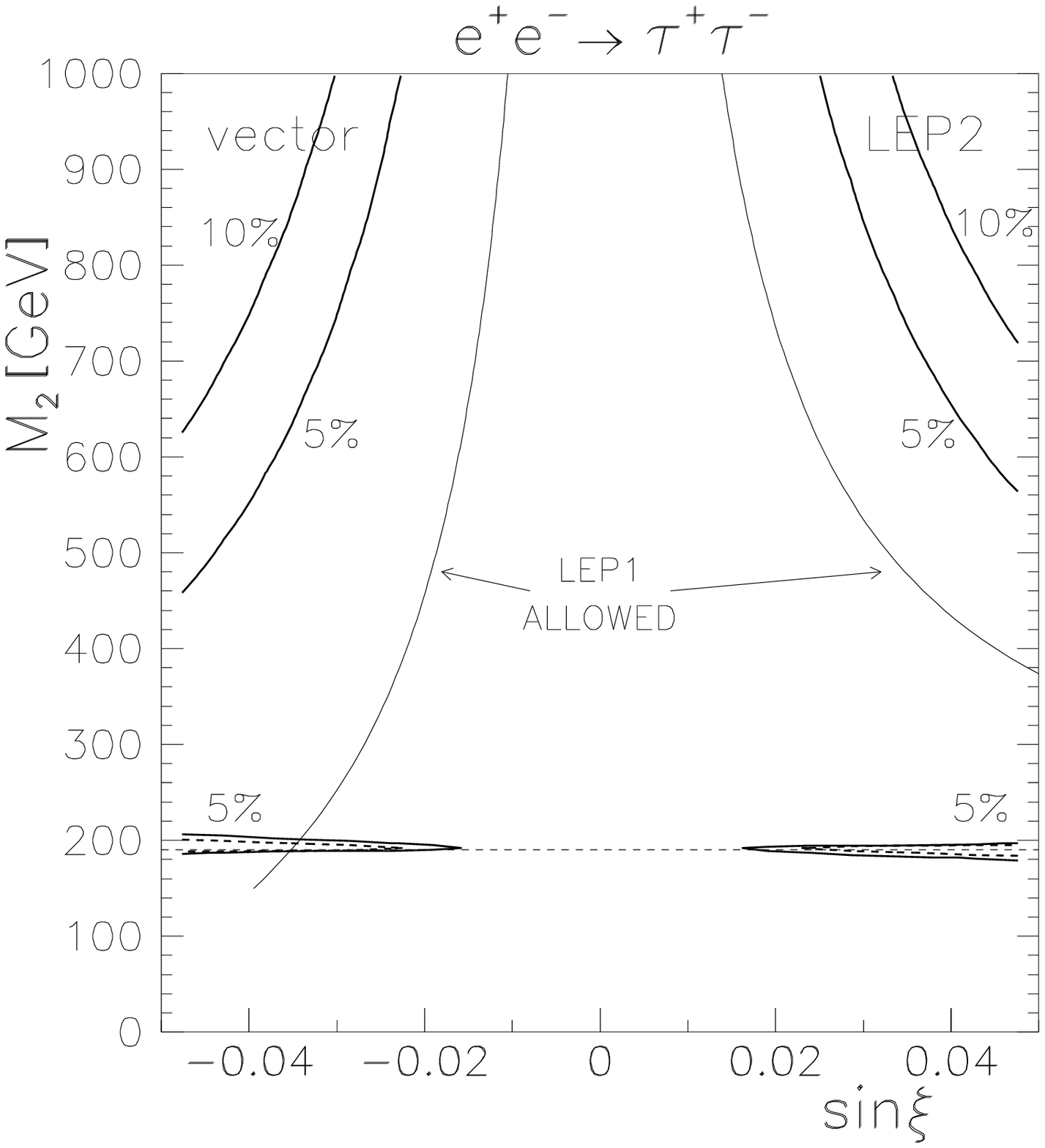}
      \epsfysize=8.5cm\epsffile{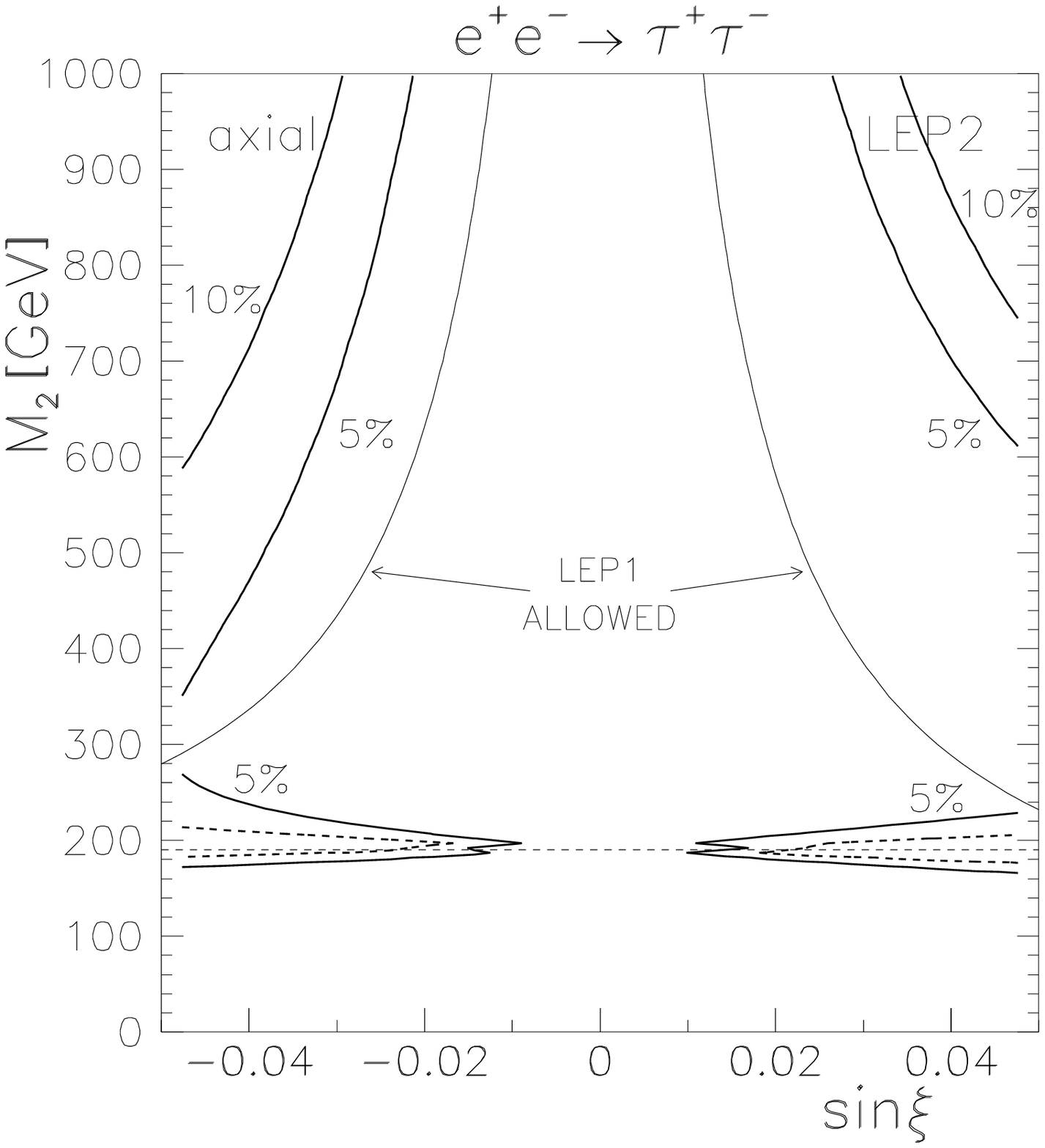}}}
\put(-1.,0.0){
\mbox{\epsfysize=8.5cm\epsffile{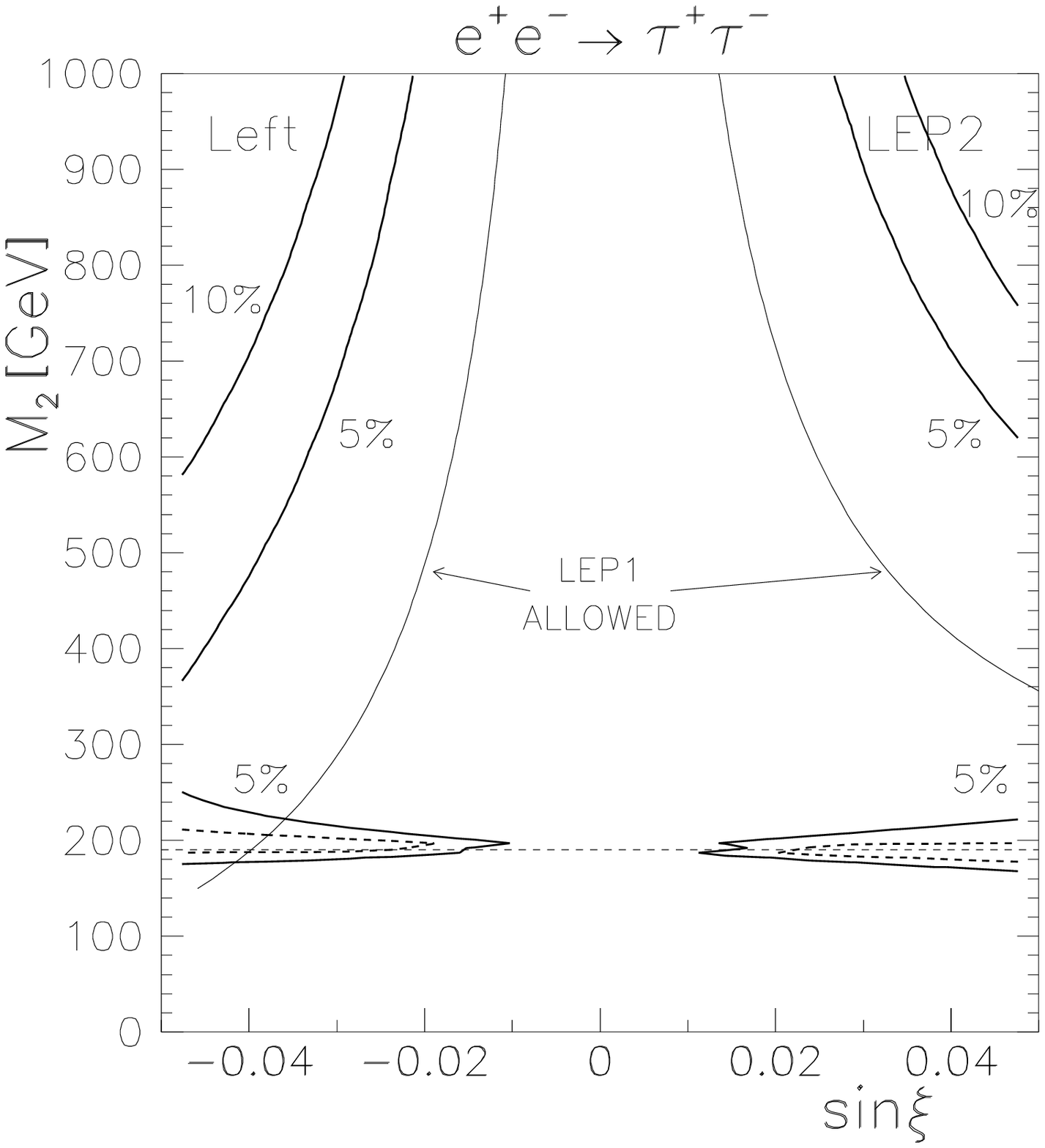}
      \epsfysize=8.5cm\epsffile{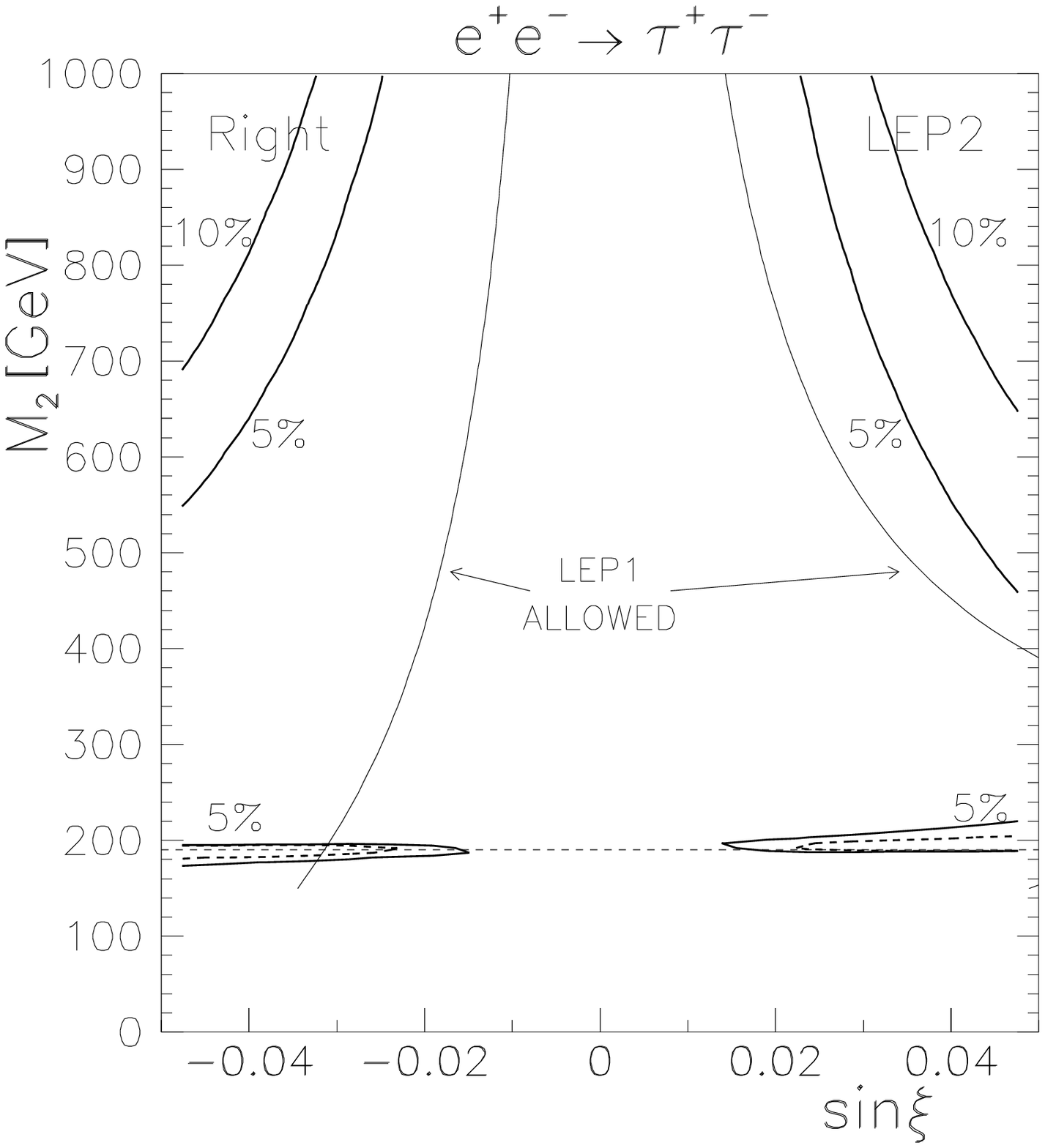}}}
\end{picture}
\begin{capt}
Allowed regions of $\sin\xi$ and $M_2$ obtained from LEP1 data
(95\% C.L.) for the process $e^+e^-\to \tau^+\tau^-$.
Also shown are bounds anticipated from LEP2 at levels of assumed precision
as indicated by labels. 
Four different chiralities are considered.
\end{capt}
\end{center}
\end{figure}

\begin{figure}
\begin{center}
\setlength{\unitlength}{1cm}
\begin{picture}(14.3,18.0)
\put(-1.,9.0){
\mbox{\epsfysize=8.5cm\epsffile{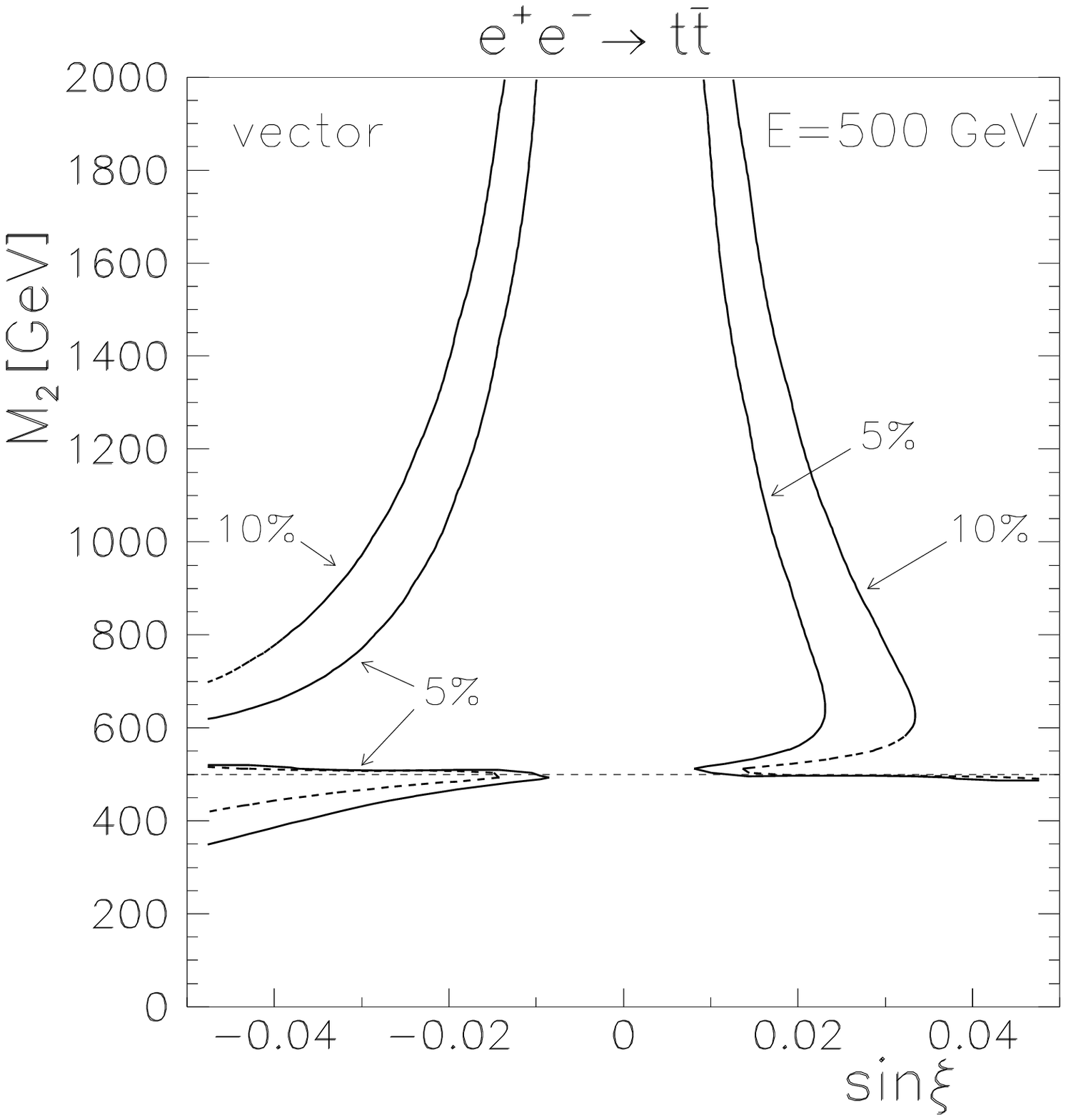}
      \epsfysize=8.5cm\epsffile{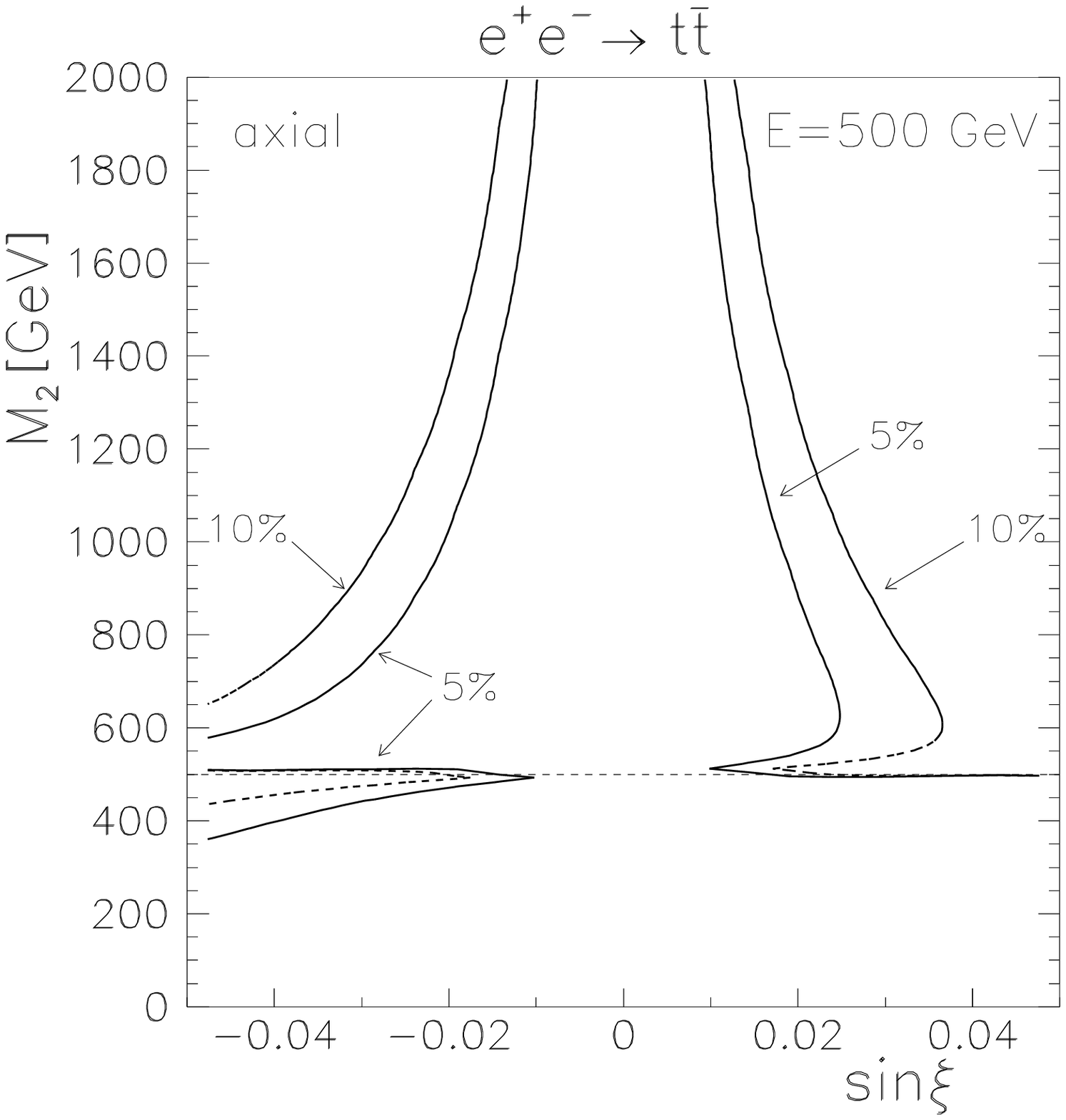}}}
\put(-1.,0.0){
\mbox{\epsfysize=8.5cm\epsffile{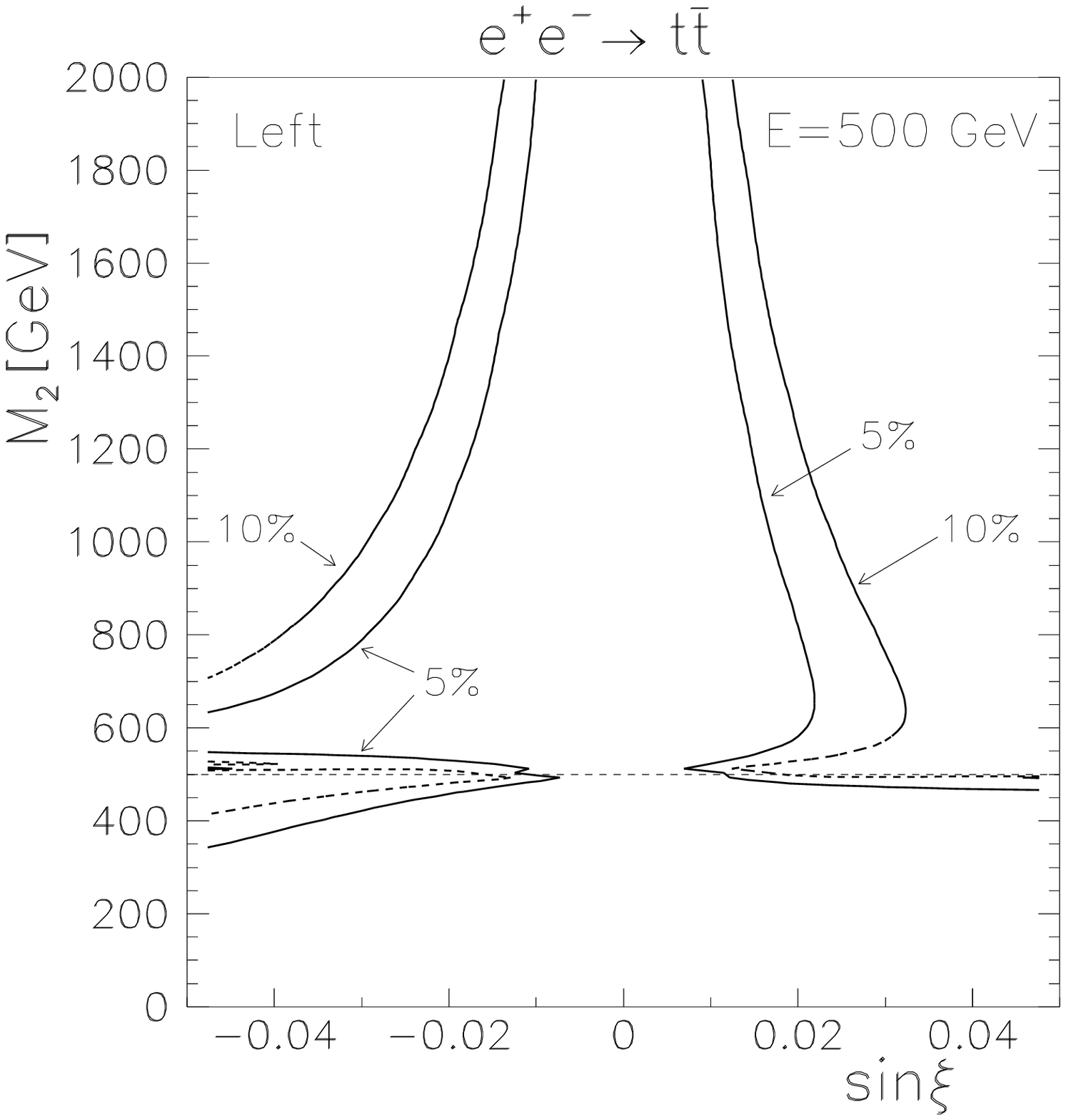}
      \epsfysize=8.5cm\epsffile{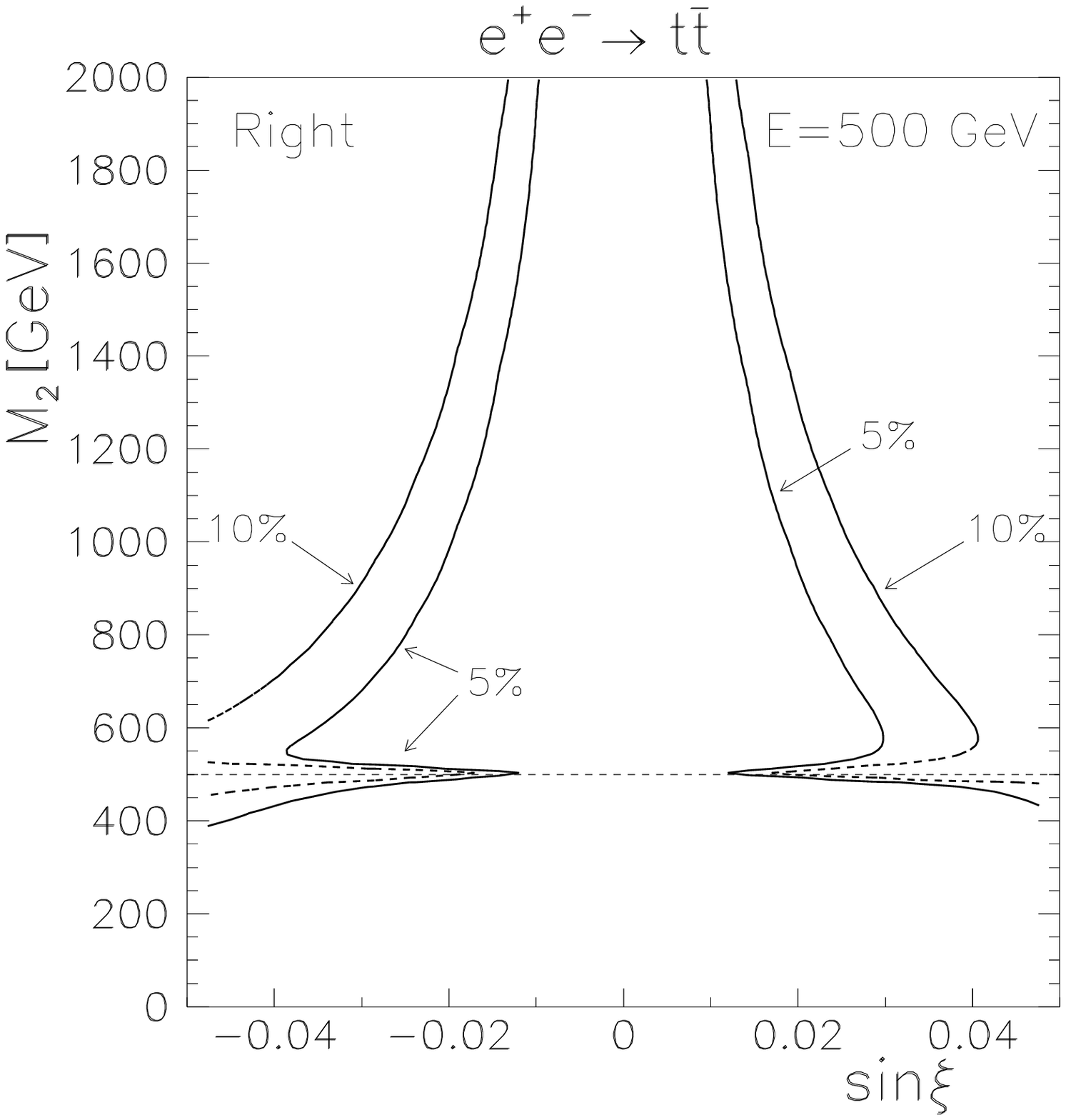}}}
\end{picture}
\begin{capt}
Allowed regions of $\sin\xi$ and $M_2$ anticipated 
for the process $e^+e^-\to t\bar t$ at the NLC
at levels of assumed precision as indicated by labels. 
Four different chiralities are considered.
\end{capt}
\end{center}
\end{figure}
\begin{figure}
\begin{center}
\setlength{\unitlength}{1cm}
\begin{picture}(14.3,18.0)
\put(-1.,0.0){
\mbox{\epsfysize=14.0cm\epsffile{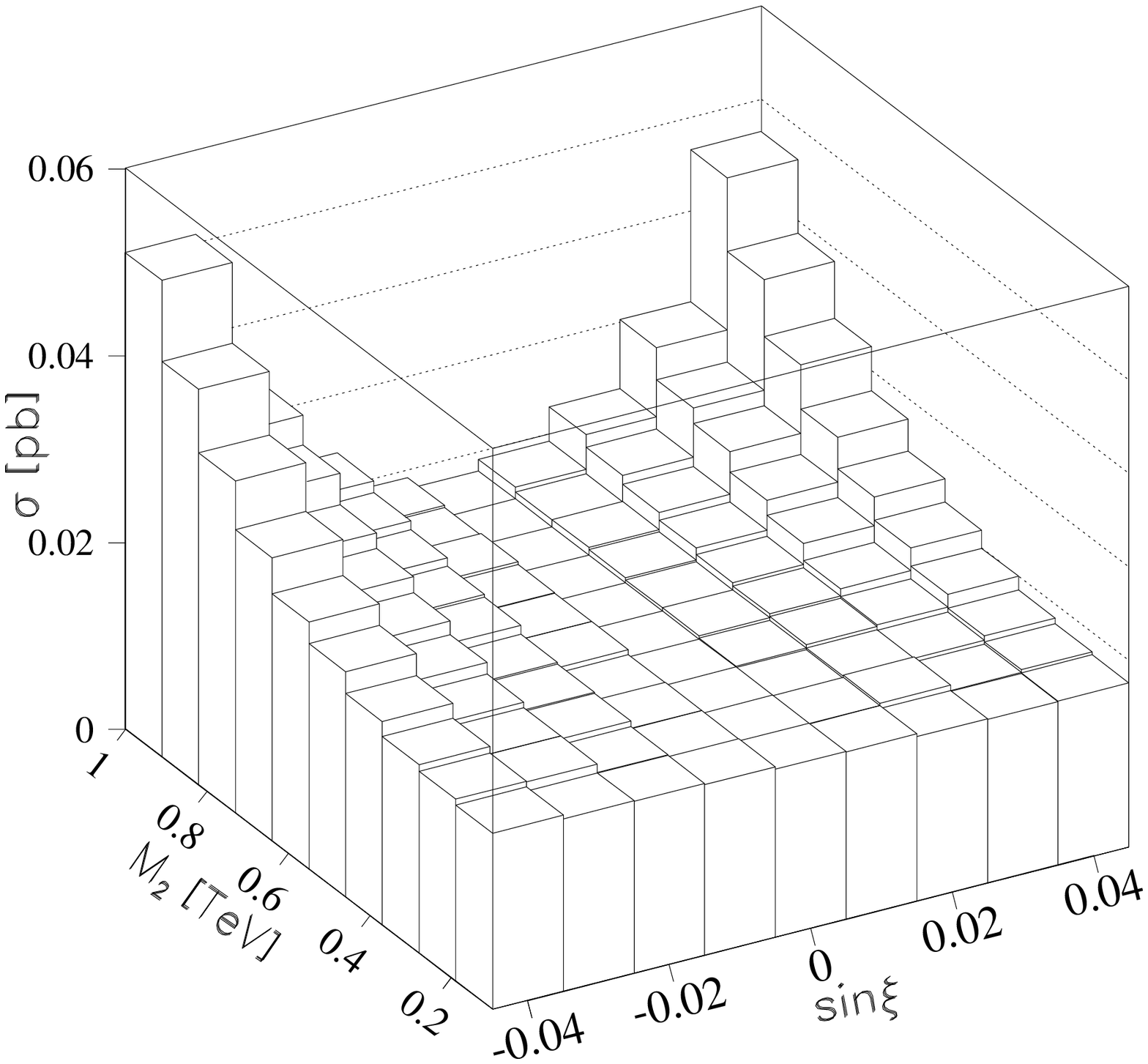}}
}
\end{picture}
\begin{capt}
Cross section for $b\bar b\nu_e \bar \nu_e$ production 
at $\sqrt{s}=190$~GeV, subject to cuts given by Eq.~(\ref{Eq:cuts}).
\end{capt}
\end{center}
\end{figure}
\begin{figure}
\begin{center}
\setlength{\unitlength}{1cm}
\begin{picture}(14.3,18.0)
\put(-1.,0.0){
\mbox{\epsfysize=14.0cm\epsffile{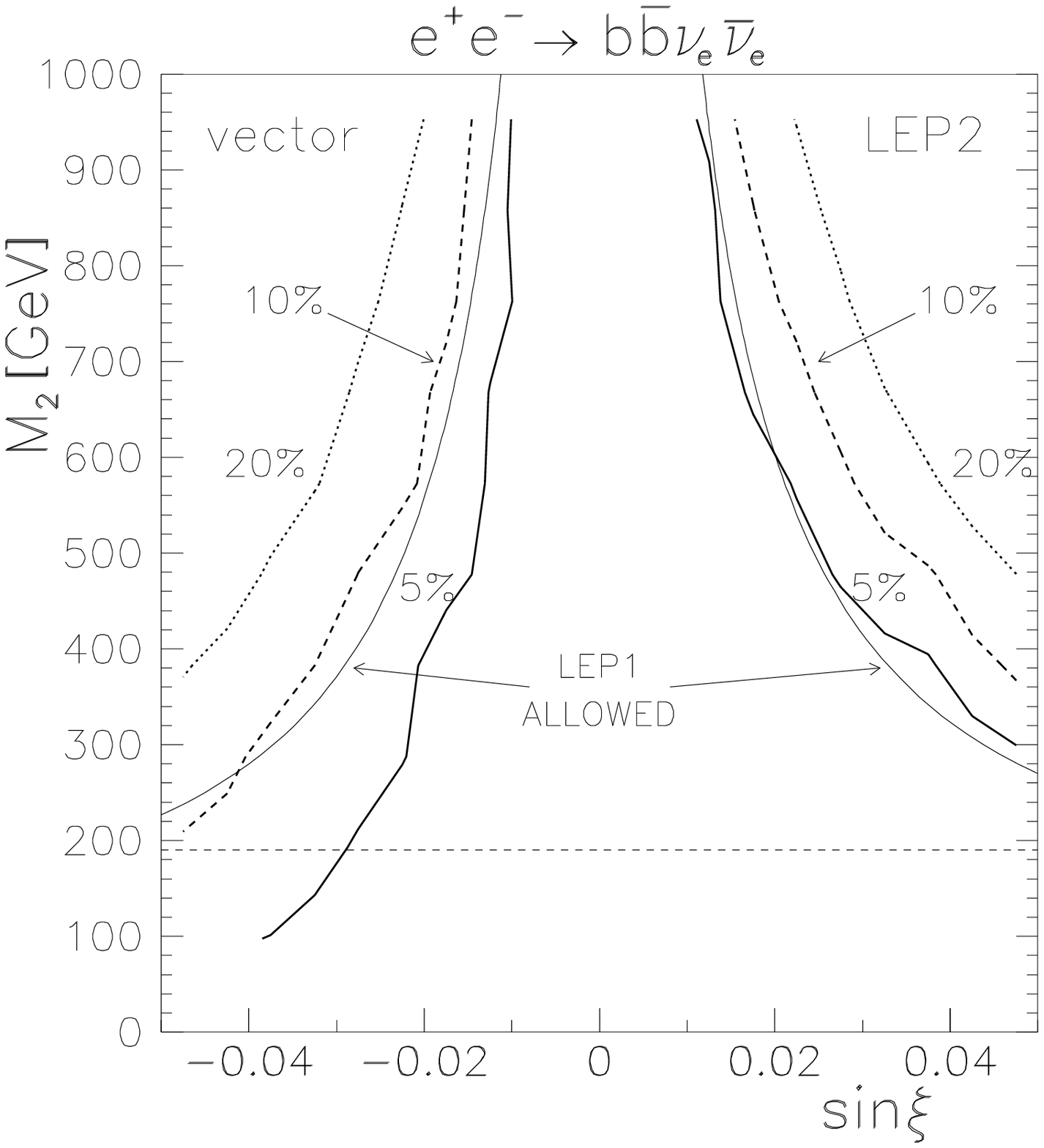}}
}
\end{picture}
\begin{capt}
Bounds anticipated from LEP2 for $e^+e^-\to b\bar b\nu_e \bar \nu_e$
at levels of assumed precision as indicated by labels.
Also shown is the
allowed region of $\sin\xi$ and $M_2$ obtained from LEP1 data
(95\% C.L.) for the process $e^+e^-\to b\bar b$.
\end{capt}
\end{center}
\end{figure}

\begin{figure}
\begin{center}
\setlength{\unitlength}{1cm}
\begin{picture}(14.3,18.0)
\put(-1.,0.0){
\mbox{\epsfysize=14.0cm\epsffile{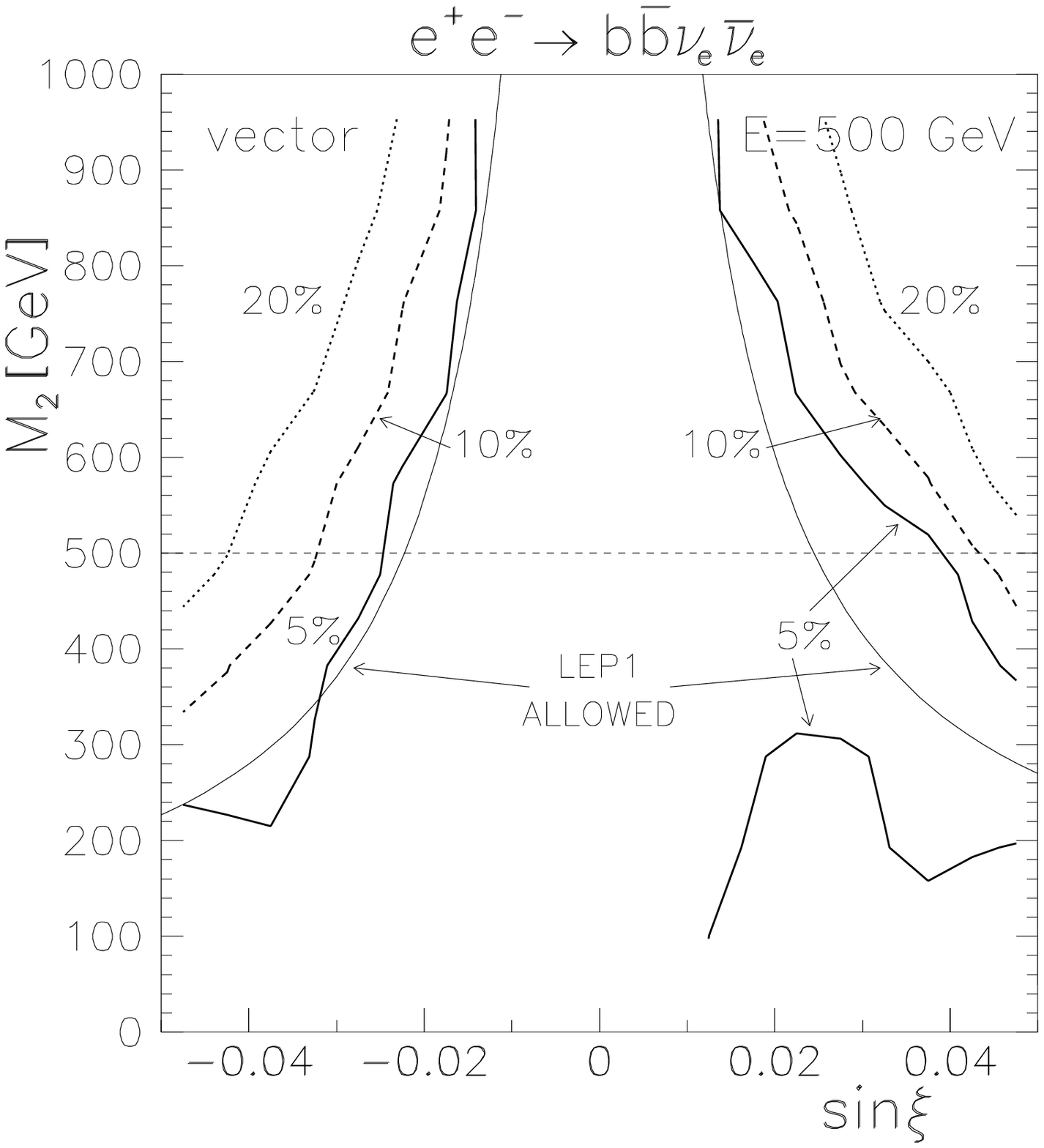}}
}
\end{picture}
\begin{capt}
Bounds anticipated from $\sqrt{s}=500$~GeV 
for $e^+e^-\to b\bar b\nu_e \bar \nu_e$
at levels of assumed precision as indicated by labels.
Also shown is the
allowed region of $\sin\xi$ and $M_2$ obtained from LEP1 data
(95\% C.L.) for the process $e^+e^-\to b\bar b$.
\end{capt}
\end{center}
\end{figure}

\end{document}